\documentclass[a4paper,12pt]{article}
\pdfoutput=1 

\usepackage{jcappub} 

\newcommand{\be}{\begin{equation}}
\newcommand{\ee}{\end{equation}}
\newcommand{\bea}{\begin{eqnarray}}
\newcommand{\eea}{\end{eqnarray}}

\newcommand{\mmu}{M_{\rm Pl}^2}
\newcommand{\eeta}{\zeta}
\newcommand{\kkappa}{\alpha}
\newcommand{\h}{y}

\newcommand{\A}{{\rm a}}

\newcommand{\vareps}{\eeta}
\newcommand{\xii}{}

\usepackage{psfrag}
\usepackage{latexsym,array,theorem,mathrsfs,subfigure,bm,float}
\usepackage{amsmath}
\usepackage{amsfonts}
\usepackage{amssymb}

\renewcommand{\theequation}{\arabic{section}.\arabic{equation}}

\title{\boldmath The screening Horndeski cosmologies}

\author[a,b]{Alexei~ A.~Starobinsky,}
\author[b]{Sergey~V.~Sushkov,}
\author[c,b,1]{Mikhail~S.~Volkov\note{Corresponding author.}}

\affiliation[a]{L.D.Landau Institute for Theoretical Physics RAS, 
Moscow 119334, Russia}

\affiliation[b]{Department of General Relativity and Gravitation, Institute of Physics,\\
Kazan Federal University, Kremlevskaya street 18, 420008 Kazan, Russia
}

\affiliation[c]{Laboratoire de Math\'{e}matiques et Physique Th\'{e}orique CNRS-UMR 7350, \\ 
Universit\'{e} de Tours, Parc de Grandmont, 37200 Tours, France}

\emailAdd{\tt alstar@landau.ac.ru}
\emailAdd{\tt sergey$_{-}$sushkov@mail.ru}
\emailAdd{\tt volkov@lmpt.univ-tours.fr}

\abstract{ 
We present a systematic  analysis of  homogeneous and isotropic cosmologies 
in a  particular Horndeski model with Galileon shift symmetry, 
containing also  a $\Lambda$-term and a matter.  
The model, sometimes called Fab Five, admits a rich spectrum of  solutions. 
Some of them describe the standard late time cosmological dynamic
 dominated by the $\Lambda$-term and matter,
while at the early times the universe expands with a
constant Hubble rate determined by the value of the scalar  kinetic coupling. For other 
solutions the $\Lambda$-term and  matter 
are screened at all times  but there are  nevertheless  the early and late accelerating  phases. 
The model also admits bounces, as well as peculiar solutions describing 
``the emergence of time''.  Most  of these solutions 
contain ghosts in the scalar and tensor sectors. However, 
a careful analysis reveals three different branches of ghost-free solutions, all 
showing a late time acceleration phase. We analyse the dynamical stability of these
solutions and find that all of them are stable in the future, since all their perturbations 
stay bounded at late times. However, they all turn out to be unstable in the 
past, as their perturbations grow  violently   when one approaches the initial spacetime 
singularity. We therefore conclude that the model has no viable solutions describing 
the whole of the cosmological history, although it may describe the current
acceleration phase. 
We also check that the flat space solution is ghost-free in the model, but it may 
acquire ghost in more general versions of the Horndeski theory. 
}

\begin{document}


\maketitle
\flushbottom

\section{Introduction -- Horndeski theory}
\setcounter{equation}{0}

The discovery of the current  universe acceleration  \cite{1538-3881-116-3-1009,0004-637X-517-2-565}
requires a theoretical explanation. From the phenomenological viewpoint,
a small cosmological term  is a very good explanation \cite{Sahni:1999gb},  
but it is problematic  from the quantum field theory viewpoint, 
since it is difficult to explain the origin and value of this term \cite{Weinberg:1988cp}. 
Therefore, alternative dark matter models have been proposed, most of which
introduce a scalar field, as in the Brans-Dicke, quintessense, $k$-essence, etc. theories (see 
\cite{Copeland:2006wr,Joyce:2014kja} for reviews), while the others, 
as for example the $F(R)$ gravity \cite{Sotiriou:2008rp,DeFelice:2010aj}, although looking different, 
are equivalent to the theory with 
a scalar field.  Some of these models were actually introduced  long ago 
in the context of the inflation theory  \cite{Starobinsky:1980te},  and some describe both 
the primordial inflation and the late time acceleration \cite{Appleby:2009uf}. 

In view of this interest towards theories with a gravitating scalar field one may ask,
what is the most general theory of this type? The answer was obtained already in 1974 by 
Horndeski  \cite{Horndeski:1974wa} --
this theory should have  at most second order field equations to avoid the Ostrogradsky ghost 
\cite{Woodard:2015zca}, 
and it is determined by the following action density (in the parameterization of   Ref.\cite{Kobayashi:2011nu}) 
\be                         
L_{\rm H}=\sqrt{-g}\,({\cal L}_2+{\cal L}_3+{\cal L}_4+{\cal L}_5), 
\ee
where, with  $X\equiv -\frac12 \nabla_\mu\Phi\nabla^\mu\Phi$, one has 
\bea                        \label{horn}
{\cal L}_2&=&G_2(X,\Phi)\,,\nonumber \\
{\cal L}_3&=&G_3(X,\Phi)\,\Box\Phi\,,\ \nonumber \\
{\cal L}_4&=&G_4(X,\Phi)\,R+\partial_X G_4(X,\Phi)\,\delta^{\mu\nu}_{\alpha\beta}\,
\nabla_\mu^\alpha\Phi\nabla_\nu^\beta\Phi\,,  \nonumber \\
{\cal L}_5&=&G_5(X,\Phi)\,G_{\mu\nu}\nabla^{\mu\nu}\Phi
-\frac16\,\partial_X G_5(X,\Phi)\,\delta^{\mu\nu\rho}_{\alpha\beta\gamma}\,
\nabla_\mu^\alpha\Phi\nabla_\nu^\beta\Phi\nabla_\rho^\gamma\Phi \,, 
\eea
the coefficient functions 
$G_k(X,\Phi)$ can be arbitrary, also
$\delta^{\lambda\rho}_{\nu\alpha}=2!\,
\delta^\lambda_{[\nu}\delta^\rho_{\alpha ]}$ and
$\delta^{\lambda\rho\sigma}_{\nu\alpha\beta}=3!\,
\delta^\lambda_{[\nu}\delta^\rho_\alpha\delta^\sigma_{\beta ]}$. 
This theory contains all previously studied models 
with a gravity-coupled scalar field.  
Recently it was rediscovered  in 
the context of the covariant Galileon models \cite{Deffayet:2009wt,Deffayet:2011gz}
(yet more recently it was found 
that it can be further 
generalised to allow  higher order derivatives in the field 
equations in such a way that the number of 
propagating degrees of freedom is still three \cite{Gleyzes:2014dya,Gleyzes:2014qga,Zumalacarregui:2013pma},
 \cite{Langlois:2015cwa,
 Crisostomi:2016tcp
 }).
Horndeski cosmologies were studied in 
Refs. \cite{Kobayashi:2011nu}, \cite{Gao:2011qe,DeFelice:2011uc,DeFelice:2011hq,DeFelice:2011bh},
\cite{Kobayashi:2015gga,Kunimitsu:2015faa,Kase:2015zva}.

The  general Horndeski theory  is 
difficult to analyse  without specifying somehow the coefficient 
functions $G_k(X,\Phi)$. There is a 
special subclass of the theory, sometimes called Fab Four (F4) \cite{Charmousis:2011bf,Charmousis:2011ea},
for which 
 the coefficients are chosen such that 
the Lagrangian becomes
\be                          \label{F40}
L_{\rm F4}=\sqrt{-g}\,({\cal L}_J+{\cal L}_P+{\cal L}_G+{\cal L}_R-2\Lambda)
\ee
with
\bea                   \label{F4}
{\cal L}_J&=&V_J(\Phi)\,G_{\mu\nu}\nabla^\mu\Phi\nabla^\nu\Phi\,,\nonumber \\
{\cal L}_P&=&V_P(\Phi)\,P_{\mu\nu\rho\sigma}\nabla^\mu\Phi\nabla^\rho\Phi\nabla^{\nu\sigma}\Phi \,,\nonumber \\
{\cal L}_G&=&V_G(\Phi)\,R \,,\nonumber \\
{\cal L}_R&=&V_R(\Phi)\,(R_{\mu\nu\alpha\beta}R^{\mu\nu\alpha\beta}
-4R_{\mu\nu}R^{\mu\nu}+R^2).
\eea
Here 
 the double dual of the Riemann tensor is 
\be
P^{\mu\nu}_{~~~\alpha\beta}=-\frac14\,\delta^{\mu\nu\gamma\delta}_{\sigma\lambda\alpha\beta}\,
R^{\sigma\lambda}_{~~~\gamma\delta}
=-R^{\mu\nu}_{~~~\alpha\beta}
+2R^\mu_{[\alpha}\delta^\nu_{\beta ]}
-2R^\nu_{[\alpha}\delta^\mu_{\beta ]}
-R\delta^\mu_{[\alpha}\delta^\nu_{\beta ]}\,,
\ee
whose contraction is the Einstein tensor, $P^{\mu\alpha}_{~~~\nu\alpha}=G^\mu_{~\nu}$.
This model is distinguished by the {\it screening property} -- it is the most general subclass 
of the Horndeski theory in which
flat space is a solution,  despite the presence 
of the cosmological term $\Lambda$. This property suggests that 
$\Lambda$ is actually irrelevant and hence there is no need to explain its value. Indeed, 
however large $\Lambda$ is, Minkowski space is always a solution and so one may hope that a
slowly accelerating universe will be a solution as well. 
Although Refs.\cite{Charmousis:2011bf,Charmousis:2011ea} did not  explain how to produce
a small value of the actual Hubble parameter,  the idea apparently was that a 
possibility for this 
should exist and should not not depend on $\Lambda$, since the latter can be screened altogether. 
The F4 cosmologies  were studied in \cite{Copeland:2012qf}, 
and it was found that the coefficient functions $V_J,\ldots ,V_R$ 
can be adjusted in such a way that the theory mimics 
all phases of the universe expansion.

A  particular model related to the F4 theories 
has received a lot of attention. 
 Setting the potential functions to constant values, $V_J=-\kkappa$, $V_P=V_R=0$, $V_G=\mmu$,
but adding  an extra term -- 
the standard kinetic term $X$ for the  scalar, 
gives  a model sometimes called Fab Five (F5) \cite{Appleby:2012rx},
\be                              \label{Fab5}
S=\frac12\int\left(\mmu\, R-(\kkappa\, G_{\mu\nu}
+\varepsilon\, g_{\mu\nu})\nabla^\mu\Phi\nabla^\nu\Phi-2\Lambda\right)\sqrt{-g}\, d^4x+S_{\rm m}
\equiv\frac12 \int L\,d^4x+S_{\rm m}\,. 
\ee
Here $M_{\rm Pl}=\sqrt{1/8\pi G}$ is the Planck mass, $\varepsilon$ is a parameter, and 
$S_{\rm m}$ describes an ordinary matter
assumed to be a perfect fluid. This model
can be integrated completely in the static and spherically symmetric sector 
\cite{Rinaldi:2012vy,Minamitsuji:2013ura,Anabalon:2013oea,Babichev:2013cya}.
Moreover, for $\varepsilon\neq 0$ it admits solutions for which 
the $\Lambda$-term is totally screened as in the F4 theory, but the metric is not flat 
but rather de Sitter with the Hubble rate proportional to $\varepsilon/\kkappa$ 
 \cite{Sushkov:2009hk,Saridakis:2010mf}, \cite{Appleby:2012rx}. 
 
 We find this model  interesting because it offers an  opportunity to describe the late time cosmic 
 self-acceleration while screening the $\Lambda$-term and hence circumventing the cosmological 
 constant problem. One should say that this model is certainly not the most general one in this respect,
 as there are other models that can also show the self-acceleration and screening. For example,
 these can be models  obtained by adding the $X$ kinetic term  to the generic F4 \eqref{F40}. 
 One can also directly modify the mini-superspace Lagrangian  such that 
 the theory admits a de Sitter solution while screening the $\Lambda$-term
 \cite{Martin-Moruno:2015bda,Martin-Moruno:2015lha}. However, we prefer to consider 
 the model \eqref{Fab5} because it is manifestly covariant, and also because it is simple enough 
 to be integrated completely. At the same time, the results we obtain suggest that the model 
 should probably be generalized to have more realistic solutions, but this can be achieved only at the sake of 
 loosing simplicity.

 In what follows we systematically study the homogeneous and isotropic cosmologies 
in the F5 model \eqref{Fab5} and we find  a rich spectrum of solutions. 
Some of them describe the standard late time cosmological dynamic
 dominated by the $\Lambda$-term and matter,
while at  early times the universe expands with a
constant Hubble rate determined by the value of the scalar  kinetic coupling. For other 
solutions the $\Lambda$-term and  matter 
are screened at all times  but there are  nevertheless  the early and late accelerating  stages. 
The model also admits bounces, as well as peculiar solutions describing a creation of universe 
``out of nothing''.  Most  of these solutions 
contain ghosts in the scalar and tensor sectors, but 
for $\varepsilon\geq 0$ and  for $\kkappa\geq 0$ there are ghost-free solutions. 
We find three different branches of such solutions, all 
showing a late time acceleration phase, and for a certain range of the 
parameter $\varepsilon$  the late time 
Hubble rate being determined by the ratio $\varepsilon/\kkappa$ and not by $\Lambda$.  
Therefore, the screening mechanism works indeed, and it is probably easier 
 to explain a small value of $\varepsilon/\kkappa$ rather than that of  $\Lambda$. 
 We also check that the flat space solution is ghost-free in the model, but it may 
acquire ghost within the full F4 theory.  

We  conclude that the model may indeed be successful, in particular at late times. 
However, it cannot apply at all times, since its radiation-dominated  solution has ghost 
and should be excluded from consideration, whereas ghost-free solutions do not show a 
radiation-dominated phase since they screen the matter together with the $\Lambda$-term.
Without this phase the model cannot correctly describe the primary nucleo-synthesis. As a result, 
the screening works ``too well" in the model. 

We also analyse the dynamical stability of the ghost-free 
solutions and find that all of them are stable in the future, since all their perturbations 
stay bounded at late times. However, they all turn out to be unstable in the 
past, as their perturbations grow  violently   when one approaches the initial spacetime 
singularity. We therefore conclude that the F5 model has no viable solutions describing 
the whole of the cosmological history. However, since it admits stable in the future solutions, 
it may well describe the current acceleration phase, hence it fulfills 
the main motivation for considering models with scalar field.  
More realistic models may probably exist in more general versions of the Horndeski theory.

The rest of  the text is organised as follows. Equations 
describing homogeneous and isotropic cosmologies are derived in the next section.
Solutions of these equations are 
constructed in Sec. III first in the early and late time limits and then globally. 
All solutions are ghost-checked and classified accordingly. Sec. IV contains the stability 
analysis of the ghost-free solutions and 
concluding remarks. Many technical details are given in the three Appendices --
the derivation of the no-ghost conditions 
in Appendix A, the  no-ghost conditions for flat space within the full F4 theory 
in Appendix B, and the equations for generic perturbations of 
spatially flat cosmologies in Appendix C. 

\section{Homogeneous and isotropic cosmologies}
\setcounter{equation}{0}

The first variation of the action \eqref{Fab5} is 
\be                                    \label{dS}
\delta S=\frac12 \int ( E_{\mu\nu}\,\delta g^{\mu\nu}+E_\Phi\, \delta \Phi)\sqrt{-g}\,d^4 x,
\ee
whose vanishing implies the gravitational equations, 
\bea                     \label{Eeq}
E_{\mu\nu}\equiv \mmu\, G_{\mu\nu}+\Lambda g_{\mu\nu}-\kkappa \,{\cal T}_{\mu\nu}-\varepsilon\, T^{(\Phi)}_{\mu\nu}-T^{\rm (m)}_{\mu\nu}=0, 
\eea
with 
\bea                       \label{TTT} 
{\cal T}_{\mu\nu}&=& P_{\alpha\mu\nu\beta}\nabla^\alpha\Phi\nabla^\beta\Phi
+\frac12\,g_{\mu\lambda}\,\delta^{\lambda\rho\sigma}_{\nu\alpha\beta}\,\nabla^\alpha_\rho\Phi\nabla^\beta_\sigma\Phi
-XG_{\mu\nu}  \,, \nonumber \\
T^{(\Phi)}_{\mu\nu}&=&\nabla_\mu\Phi\nabla_\nu\Phi+Xg_{\mu\nu} \,,     \nonumber \\
T^{\rm (m)}_{\mu\nu}&=&(\rho+p)U_\mu U_\mu+pg_{\mu\nu}\,,
\eea
and  the scalar equation 
\be                 \label{scal}
E_\Phi\equiv 
\nabla_\mu((\kkappa G^{\mu\nu}+\varepsilon g^{\mu\nu})\nabla_\nu\Phi)=0\,. 
\ee
This latter equation has the structure of current conservation, due to the theory 
 invariance under shifts   $\Phi\to\Phi+\Phi_0$.
The structure of this equation also implies that the propagation of $\Phi$
is determined by the effective ``optical" metric ${\cal M}_{\mu\nu}=\kkappa G_{\mu\nu}+\varepsilon g_{\mu\nu}$.  
Since the energy-momentum tensors \eqref{TTT} are obtained by varying  
the diffeomorphism-invariant pieces of the action, each of them is independently conserved, 
 hence one has on-shell $\nabla^\mu {\cal T}_{\mu\nu}=0$, 
$\nabla^\mu T^{(\Phi)}_{\mu\nu}=0$, $\nabla^\mu T^{\rm (m)}_{\mu\nu}=0$.

Let us choose the FLRW ansatz  for the metric,  
\be                            \label{FLRW}
ds^2=-dt^2+\A^2(t)\left[
\frac{dr^2}{1-Kr^2}+r^2 (d\vartheta^2+\sin^2\vartheta d\varphi^2)
\right],
\ee
with $K=0,\pm 1$. Denoting $H=\dot{\A}/\A$ the Hubble parameter and 
$\psi=\dot{\Phi}$, 
the non-trivial gravitational equations \eqref{Eeq} are 
\bea             \label{eq}
E_0^0&\equiv& -3\mmu\left(H^2+\frac{K}{\A^2}\right)+
\frac{1}{2}\,\varepsilon\,\psi^2-\frac{3}{2}\,\kkappa\,\psi^2\left(3H^2+\frac{K}{\A^2}\right)
+\Lambda+\rho=0,  \\
E_1^1&\equiv& -\mmu\left(2\dot{H}+3H^2+\frac{K}{\A^2}
\right)-\frac12\,\varepsilon\,\psi^2    \nonumber \\
&-&\kkappa\,\psi^2\left(
\dot{H}+\frac32\,H^2-\frac{K}{\A^2}+2H\frac{\dot{\psi}}{\psi}
\right)+\Lambda-p=0, 
\nonumber 
\eea
and also
$E^1_1=E^2_2=E^3_3$, 
while 
the scalar field equation \eqref{scal} is 
\be                            \label{eqq}
E_\Phi\equiv \frac{1}{\A^3}\frac{d}{dt}\left(
\A^3\left(3\kkappa\,\left(H^2+\frac{K}{\A^2}\right)-\varepsilon\right)\psi
\right)=0. 
\ee
It is straightforward to check that 
\be                            \label{Bian}
\dot{E}_0^0+3H(E_0^0-E_1^1)+E_\Phi=0,
\ee
in view of the matter conservation condition $\dot{\rho}+3H(\rho+p)=0$,
hence only two of the three equations \eqref{eq},\eqref{eqq} are independent. 
(These equations were recently applied to study a homogeneous collapse
of the FRLW metric \eqref{FLRW} with $K=+1$ matched to an exterior vacuum
space-time to describe  black hole formation from the point of view
of an external observer \cite{Koutsoumbas:2015ekk}.)

The first integral of the scalar field equation \eqref{eqq} is 
\bea 										 \label{eq1}
\A^3\left(3\kkappa\,\left(H^2+\frac{K}{\A^2}\right)-\varepsilon\right)\psi=C,
\eea
where $C$ is the Noether charge associated with the shift symmetry $\Phi\to\Phi+\Phi_0$. 
Let us first set  $C=0$. One finds in this case 
 two different solutions  
which we shall call
GR branch and screening branch. 
It turns out that solutions with $C\neq 0$ always approach one of these 
branches at late times.   

\underline{The GR branch} is obtained by setting $\psi=0$. This solves the scalar equation \eqref{eq1}, 
while the gravitational equation \eqref{eq} reduces  to 
\be                             \label{GR}
H^2+\frac{K}{\A^2}=\frac{\Lambda+\rho}{3\mmu}. 
\ee
Since the matter density $\rho$ tends to zero at late time, the expansion is driven by $\Lambda$. 

\underline{The screening branch} is obtained by setting to zero the expression in the  parenthesis in \eqref{eq1},
\be                             \label{Fab}
H^2+\frac{K}{\A^2}=\frac{\varepsilon}{3\kkappa}\,.
\ee
This solves the scalar field equation, but the solution determines the metric and not the scalar field. 
The latter is determined by the gravitational equation \eqref{eq},
\be                            \label{Fab1}
\psi^2=\frac{\kkappa\,(\Lambda+\rho)-\varepsilon\,\mmu}{\kkappa\,\left(\varepsilon -3\kkappa\,{K}/{\A^2}\right)}. 
\ee
The role of the cosmological constant is now played  by $\varepsilon/3\kkappa$ while the $\Lambda$-term
is screened and makes no contribution  to the  universe acceleration. 
Note that the matter density $\rho$ is screened in the same sense, too.
This applies for all (spatially open, closed, and flat) types of solutions, hence
the spacetime is that of a constant curvature, de Sitter or
anti-de Sitter, depending on the relative sign of $\varepsilon$
and $\kkappa$ (we shall see below that the absence of ghost requires that 
$\varepsilon\geq 0$ and $\kkappa\geq 0$).  

If $\varepsilon=0$ then the F5 theory becomes F4 and the following 
{\it flat metric} configuration solves the field equations (which can be seen from Eqs.\eqref{Fab},\eqref{Fab1}),
\be                            \label{Fab2}
K=-1,~~~~~  \A=t,~~~~~~\psi^2=\frac{\Lambda+\rho}{3\kkappa}\,t^2\,.
\ee
This is the principal virtue  of the F4 theory -- to admit 
a flat  solution despite the non-zero $\Lambda$,  hence the cosmological term 
can be {\it totally screened} (as well as $\rho$) 
\cite{Charmousis:2011bf,Charmousis:2011ea}. 

Let us now see what happens if  $C\neq 0$. Eq.\eqref{eq1} then yields
\be                                \label{psi}
\psi=\frac{C}{\A^3\,\left[3\kkappa\, (H^2+\frac{K}{\A^2})-\varepsilon\right]},
\ee
injecting which to \eqref{eq} gives 
\be                   \label{HH}
3\mmu\left(H^2+\frac{K}{\A^2}\right)=
\frac{C^2\left[\varepsilon-3\kkappa\left(3 H^2+ \frac{K}{\A^2}\right)\right]}
{2a^6\, \left[\varepsilon-3\kkappa \,(H^2+\frac{K}{\A^2})\right]^2   }+\Lambda+\rho. 
\ee
This equation determines the algebraic dependence of the Hubble parameter $H$ on the scale factor $\A$.
The relation to the physical time is then determined by the quadrature
\be
t=\int \frac{d\A}{\A H(\A)}.
\ee
Let us 
set 
\be    							\label{scale0}
H^2=H_0^2\, \h,~~~~~\A=\A_0\, a\,,~~~~~~\rho_{\rm cr}=3\mmu H_0^2\,,
\ee
where $H_0$ and $\A_0$ are the actual values of the Hubble parameter and of the scale factor, whereas 
$\rho_{\rm cr}$ is the critical density. 
Assuming the matter to be a mixture of  a radiation and a non-relativistic component, 
\be
\rho=\rho_{\rm cr}\left(\frac{\Omega_4}{a^4}+\frac{\Omega_3}{a^3}\right),
\ee
Eq.\eqref{HH} becomes 
\be                          \label{HHH}
\h=\Omega_0+\frac{\Omega_2}{a^2}+\frac{\Omega_3}{a^3}+\frac{\Omega_4}{a^4}
+\frac{\Omega_6 \left[\vareps-3\h+\frac{\Omega_2}{a^2}  \right]}{a^6 \left[\vareps-\h+\frac{\Omega_2}{a^2}  \right] ^2},
\ee
where
\be                       \label{Om6}
\Omega_0=\frac{\Lambda}{\rho_{\rm cr}},~~~~\Omega_2=-\frac{K}{H_0^2 \A_0^2},~~~~~~
\Omega_6=\frac{C^2}{6\kkappa\,\A_0^6\,H_0^2\,\rho_{\rm cr}},~~~~~\vareps=\frac{\varepsilon}{3\kkappa\,H_0^2}\,.
\ee
We assume in this text that $\Lambda>0$, hence $\Omega_0$ is always positive. The sign of $\Omega_2$
is opposite to that of $K$, while the sign of $\Omega_6$ is the same as that of $\kkappa$.

The scalar field $\psi$ in \eqref{psi} can be expressed in terms of a dimensionless  $\Psi$ as 
\be                   \label{PSI}
\psi=-\sqrt{\frac{2\rho_{\rm cr}\Omega_6 }{3\kkappa H_0^2 } }\,\Psi~~~~~~~{\rm where}~
~~~~~~\Psi=\frac{1}{a\,[a^2(\eeta-\h)+\Omega_2] }\,.
\ee
Before studying Eq.\eqref{HHH}, let us introduce the conditions for the absence of ghosts derived in 
Appendix A. The ghosts may arise for generic perturbations of solutions of \eqref{HHH}.
However, they will be absent for the tensor and vector perturbations if $\kkappa\geq 0$, hence if 
\be                          \label{gh1}
\Omega_6\geq 0,
\ee
while their absence in the scalar sector  requires that
\be                           \label{gh2}
{\cal G}\equiv \left[\h\,\left(\h_\ast-\h\right)^2a^6
+\Omega_6\,\left(6\h-\h_\ast\right)\right]\left[(\h_\ast-\h)^3a^6+\Omega_6\,(3\h+\h_\ast)\right]>0, 
\ee
where 
\be                                       \label{hstar}
\h_\ast={\vareps}+\frac{\Omega_2}{a^2}.
\ee
Therefore, our goal is to study solutions of the algebraic equation 
\eqref{HHH} subject to conditions \eqref{gh1} and \eqref{gh2}. 
Bringing all terms in \eqref{HHH} to the common denominator yields 
\be                             \label{NN} 
\frac{P(a,\h)}{a^8\,[\h-\h_\ast]^2}=0,
\ee
where 
\be       \label{N}
P(a,\h)=c_3(a)\, \h^3+c_2(a)\, \h^2+c_1(a)\, \h +c_0(a)
\ee
is the cubic in $\h$ polynomial with the coefficients 
\bea
c_3&=&-a^8\,,~~~~
c_2=(\Omega_2+2\eeta)\,a^8+3\Omega_2 \,a^6+\Omega_3 \,a^5+\Omega_4\, a^4\,, ~~~
c_1=-\eeta(2\Omega_0+\eeta)\,a^8 \nonumber \\
&-&2\Omega_2(\Omega_0+2\eeta)\,a^6-2\eeta\Omega_3\,a^5-(3\Omega_2^2+2\eeta\Omega_4)\,a^4 
-2\Omega_2\Omega_3\, a^3-(2\Omega_2\Omega_4+3\Omega_6)\,a^2\,,  \nonumber \\
c_0&=&\eeta^2\Omega_0\,a^8+\eeta\Omega_2(2\Omega_0+\eeta)\,a^6+\eeta^2\Omega_3\, a^5
+(\Omega_4\eeta^2+\Omega_2^2(\Omega_0+2\eeta))\,a^4 +2\eeta \Omega_2\Omega_3\, a^3\nonumber \\
&+&(\Omega_2^3+2\eeta\Omega_2\Omega_4+\eeta\Omega_6)\,a^2
+\Omega_2^2\Omega_3 \,a+\Omega_2(\Omega_2\Omega_4+\Omega_6). 
\eea
Eq.\eqref{NN} will be fulfilled if $P(a,\h)=0$ and $\h\neq \h_\ast$, hence  the problem 
reduces to studying roots of the cubic polynomial. We notice that 
a cubic polynomial always has one real root, and it will have two more real roots if 
its coefficients fulfill the following two conditions, 
\bea                   \label{cond}
\Delta&=&(c_2)^2-3\,c_1c_3>0,  \nonumber \\
D&=&27\, (c_0c_3)^2-18\, c_0c_1c_2c_3+4\,c_0(c_2)^3+4\,(c_1)^3c_3-(c_1c_2)^2<0.
\eea 
These conditions insure that the polynomial 
has two real extrema of the opposite sign.

\section{Constructing the solutions}
\setcounter{equation}{0}

Our  next task is to solve Eq.\eqref{HHH} (see 
\cite{Sushkov:2012} for the $\Omega_2=\Omega_4=0$ case), for which we shall first analyse
the limits where the scale factor $a$ is either large or small, and then construct the solutions globally. 

\subsection{Late time limit $a\to\infty $}

If $a$ is large  then there is always a solution of \eqref{HHH}
approaching the GR branch
\eqref{GR}, 
\be                                           \label{GRR}
\h=\Omega_0+\frac{\Omega_2}{a^2}+\frac{\Omega_3}{a^3}+\frac{\Omega_4}{a^4}
+\frac{(\vareps-3\xii\,\Omega_0 )\,\Omega_6}{(\xii\,\Omega_0-\vareps)^2\,a^6}
+{\cal O}\left(\frac{1}{a^7}\right),
\ee
the corresponding dimensionless scalar  being
\be                                         \label{GRR1}
\Psi=\frac{1}{(\eeta-\Omega_0)\,a^3}+{\cal O}\left(\frac{1}{a^6}\right). 
\ee
The ghost function ${\cal G}$   \eqref{gh2} reduces in the leading order to 
\be                                         \label{GRR2}
{\cal G}=\Omega_0(\eeta-\Omega_0)^5 a^{12}+\ldots>0,
\ee
hence, since $\Omega_0>0$,  ghost will be absent if $\eeta>\Omega_0$. 

Next, 
Eq.\eqref{cond} yields in the leading order 
$$\Delta=(\Omega_0-\eeta)^2 a^{16}+\ldots,~~~~~~
D=-8\,\Omega_6\,\eeta\,(\Omega_0-\eeta)^3 a^{26}+\ldots,
$$
where the dots denote subleading terms. Hence, as $\Omega_6>0$,  conditions 
 \eqref{cond} are fulfilled if
\be                                               \label{c1}
\eeta\, (\xii\,\Omega_0-\vareps)>0,
\ee
in which case there are  two more solutions 
$h(a)$ at large $a$. 
They approach the screening branch \eqref{Fab} and have the structure
\be                       \label{scr}
\h_\pm={\vareps}+\frac{\Omega_2}{a^2}
\pm \frac{{\chi}}{\xii\,(\xii\,\Omega_0-\vareps)\,a^3}
\pm \frac{\xii\,\Omega_2\Omega_6 }{{\chi} \,a^5}\
-\frac{\Omega_6(\eeta-3\Omega_0)\pm \Omega_3{\chi}}{2(\Omega_0-\eeta)^2\,a^6}
+{\cal O}\left(\frac{1}{a^7}\right), 
\ee
with $\chi=\sqrt{ 2\,\vareps\,\xii\,\Omega_6(\xii\,\Omega_0-\vareps)}$, while
\be
\Psi_\pm=\mp\sqrt{\frac{\Omega_0-\eeta  }{2\eeta\, \Omega_6   }  }+{\cal O}\left(\frac{1}{a^2}\right)\,.
\ee
Calculating again the leading terms of the function ${\cal G}$ in \eqref{gh2} shows that 
both $\h_\pm$ will be ghost-free if 
\be
5+\frac{2\eeta}{\Omega_0-\eeta}>0\,.
\ee
These facts can be summarized by dividing the values of $\eeta$ into three regions as follows.

{\bf I:} $\eeta>\Omega_0$ $\Rightarrow$ only the GR solution \eqref{GRR} exists -- stable (no ghost). 

{\bf II:} $\eeta<0$ $\Rightarrow$  only the GR solution \eqref{GRR} exists -- unstable. 

{\bf III:} $0<\eeta<\Omega_0$ $\Rightarrow$ there are three solutions. The GR solution \eqref{GRR} is unstable
but the two screening solutions \eqref{scr} are  stable. 

One should emphasize that  by  ``stable solutions" we mean, in this Section only,  solutions which 
do not show the ghost instability. However, they may have other instabilities, which 
 will be discussed in the next Section.

Let us now consider the  special cases $\eeta=0$ and $\eeta=\Omega_0$. If $\eeta=\Omega_0\neq 0$ then 
there is only one solution, 
\be
\h=\eeta+\frac{\Omega_2+\xi}{a^2}+\frac{\Omega_3}{3 a^3}+{\cal O}\left(\frac{1}{a^4}\right)\,,~~~~~~~
\Psi=-\frac{1}{\xi\,a}+{\cal O}\left(\frac{1}{a^2}\right)\,,
\ee
where $\xi=(-2\eeta\,\Omega_3)^{1/3}$. This solution is stable.

For $\eeta=0$ there are three different solutions. Let us first assume that $\Omega_2\neq 0$. 
Then one solution is obtained by simply setting $\eeta=0$ in Eqs.\eqref{GRR},\eqref{GRR1},
and Eq.\eqref{GRR2} then shows that this solution is unstable. The two other solutions are 
\be                         \label{pm}
\h_\pm=\frac{\Omega_2}{a^2}\pm \frac{\xi}{a^4}+\frac{3\Omega_6}{2\Omega_0\, a^6}
\mp\frac{\xi\Omega_3}{2\Omega_0\,a^7}
+{\cal O}\left(\frac{1}{a^8}\right)\,,
~~~~~~\Psi_\pm=\mp\frac{a}{\xi}+{\cal O}\left(\frac{1}{a}\right),
\ee
where $\xi=\sqrt{2\Omega_2\Omega_6/\Omega_0}$.
These solutions are stable, since 
$
{\cal G}=20\Omega_2^2\Omega_6^2/a^4+\ldots
$
Both of them approach  flat geometry if $\Omega_2>0$ ($K=-1$).

 If $\Omega_2=\eeta=0$ then solutions 
 can be obtained analytically, these are $\h=0$ and 
 \be                                 \label{eta0}
 \h_\pm=\frac12\left(\Omega_0+\frac{\Omega_3}{a^3}+\frac{\Omega_4}{a^4}\pm
 \sqrt{
 \left(\Omega_0+\frac{\Omega_3}{a^3}+\frac{\Omega_4}{a^4}\right)^2-\frac{12\,\Omega_6}{a^6}
 }
 \right).
 \ee
 In the latter case Eq.\eqref{gh2} gives 
 $
{\cal G}=\h^2(\h^2 a^6+6\Omega_6)(3 \Omega_6-\h^2 a^6). 
 $
The  $\h=0$ configuration is actually an artefact and not a solution 
of the original problem because $\psi$ in \eqref{PSI}
 is not defined if $\eeta=\h=\Omega_2=0$. Next, one has 
 $\h_{+}\to \Omega_0$ and  hence ${\cal G}\to -\Omega^6_0\, a^{12}$ at large $a$, therefore this 
 solution is unstable. The $\h_{-}$ solution has the following behaviour at large and small $a$, 
 \be
 \frac{3\Omega_6}{(\Omega_0 a^4+\Omega_3 a+\Omega_4)a^2}
 \leftarrow \h_{-}\to\frac{3\Omega_6}{\Omega_0\, a^6}, 
 \ee 
 hence ${\cal G}$ is positive at large $a$ but it can be negative at small $a$, for example
 if $\Omega_3=\Omega_4=0$.

 The above analysis exhausts all possible types of the late time solutions. 
 However, for $\eeta=0$  there is one more solution -- 
 flat space   described by \eqref{Fab2}.
 This solution should be considered separately, 
 because it has $C=\Omega_6=0$ but $\psi\neq 0$. 
 Inserting this solution to Eq.\eqref{g-s1} in the Appendix A gives 
 the positive eigenvalue of the kinetic energy matrix, 
  \be
 \lambda_1=t\,(2\mmu+5\kkappa\psi^2)>0,
 \ee
 which implies that Minkowski space in ghost-free.
 However, it may be unstable  within the full F4 theory -- the corresponding 
 no-ghost conditions are given in Appendix B.
 
 Summarizing, stable late time solutions
for generic values of $\kkappa$  are either 
 \be                    \label{infty1}
 \h=\Omega_0+\ldots,~~~~~\Psi=\frac{1}{(\kkappa-\Omega_0)\,a^3}+\ldots 
 \ee
 if $\kkappa>\Omega_0$ or 
 \be                   \label{infty2}
 \h=\eeta+\ldots,~~~~~
\Psi=\pm\sqrt{\frac{\Omega_0-\eeta  }{2\eeta\, \Omega_6   }  }+\ldots
\ee
 if $0<\kkappa<\Omega_0$.

\subsection{Limit $a\to 0$}

Let us now consider the $a\to 0$ limit, assuming first that $\underline{\Omega_4\neq 0}$.  
Then there is always the GR type solution, 
\be                              \label{GR0}
\h=\frac{\Omega_4}{a^4}+\frac{\Omega_3}{a^3}+\frac{\Omega_2\Omega_4-3\Omega_6}{\Omega_4 a^2}
+\frac{3\Omega_3\Omega_6}{\Omega_4 a}+{\cal O}(1),~~~~~\Psi=-\frac{a}{\Omega_4}\,+{\cal O}(a^2), 
\ee
but it is unstable since ${\cal G}=-\Omega_4^6/a^{12}+\ldots$. 
Next,  computing the leading terms of the $\Delta,D$ coefficients \eqref{cond} 
one finds that  there should be two more solutions if
\be                   \label{BOT}
8\Omega_2\Omega_4+9\Omega_6>0. 
\ee
If this condition is fulfilled, then introducing 
$\sigma=\sqrt{\Omega_6(8\Omega_2\Omega_4+9\Omega_6)}$, the two other solutions read 
\be
\h_\pm =\frac{2\Omega_2\Omega_4+3\Omega_6\pm \sigma}{2\Omega_4 a^2}\mp
\frac{\Omega_3\Omega_6(4\Omega_2\Omega_4+9\Omega_6\pm 3\sigma)}{2\sigma\Omega_4^2\, a}+{\cal O}(1),
~\Psi_\pm=-\frac{2\Omega_4}{(3\Omega_6\pm\sigma)a}+{\cal O}(1),
\ee
whose stability conditions are, respectively,
\be
(5\Omega_2\Omega_4\pm 3\sigma +9\Omega_6)(8\Omega_2\Omega_4\pm 3\sigma +9\Omega_6)>0. 
\ee
These solutions can be called screening, since the matter contribution, usually dominant at small $a$,
is screened. 
If $\Omega_2=0$ then nothing special 
happens to the GR solution \eqref{GR0} and it remains unstable, while the screening 
solutions 
$\h_\pm$ become 
\bea                                           \label{Om0}
\h_{+}&=&\frac{3\Omega_6}{\Omega_4\, a^2}-\frac{3\Omega_3\Omega_6}{\Omega_4^2\,a}+\frac53\,\vareps
+\frac{3\Omega_6\Omega_3^2+9\Omega_6^2}{\Omega_4^3}+{\cal O}(a), 
~~~~~\Psi_{+}=-\frac{\Omega_4}{3\Omega_6 a}+{\cal O}(1),
\nonumber \\
\h_{-}&=&\frac{\vareps}{3\,\xii}+\frac{4\,\vareps^2}{27\,\xii\,\Omega_6}\left(\Omega_4\,a^2+\Omega_3\,a^3\right)+{\cal O}(a^4),~~~~~~~~\Psi_{-}=\frac{3}{2\eeta a^3}+{\cal O}\left(\frac{1}{a}\right), 
\eea
and these are both stable. 

Let us now set $\underline{\Omega_4=0}$. Then there is always a screening solution, 
\be                                    \label{scrn1}
\h=\frac{\Omega_2}{3a^2}+\frac{4\Omega_2^2\Omega_3}{27\Omega_6 a}+\frac{\eeta}{3}+\frac{8\Omega_2^3}{81\Omega_6}-\frac{16\Omega_2^3\Omega_3^3}{243\Omega_6^2}+{\cal O}(a),~~~~
\Psi=\frac{3}{2\Omega_2 a}+{\cal O}(1),
\ee
with ${\cal G}=2\Omega_2\Omega_6^2/a^4+\ldots$, hence it is stable. 
Computing again the leading terms of $\Delta,D$ in \eqref{cond}, one finds 
two more solutions if 
\be                         \label{con0}
\Omega_3^2-12\Omega_6>0. 
\ee
Introducing $\omega=\sqrt{\Omega_3^2-12\Omega_6}$ these solutions are 
\be                                   \label{scrn2}
\h_\pm=\frac{\Omega_3\pm\omega}{2a^3}
+\frac{\Omega_2(\Omega_3^2\pm\omega\Omega_3-16\Omega_6)}
{\omega(\omega\pm\Omega_3)a^2}+{\cal O}\left(\frac{1}{a}\right),~~~~
\Psi_\pm=-\frac{2}{\Omega_3\pm\omega}+{\cal O}(a),
\ee
with the stability condition 
\be
(\Omega_3^2\pm\omega\Omega_3+6\Omega_6)(\Omega_3^2\pm\omega\Omega_3-12\Omega_6)<0.
\ee
A simple analysis of this condition shows that $\h_{-}$ is always stable while $\h_{+}$ is always unstable. 
If $\Omega_2=0$ then \eqref{scrn1},\eqref{scrn2} reduce to 
\bea                 \label{inf1}
\h&=&\frac{\vareps}{3\,\xii}+\frac{4\,\vareps^2\Omega_3}{27\,\xii\,\Omega_6}\,a^3+{\cal O}\left(a^6\right), 
~~~~~~~~~~~~~~~\Psi=\frac{3}{2\eta a^3}+{\cal O}(1),
\nonumber \\
\h_\pm&=&\frac{\xii\,\Omega_3\pm{\omega}}{2\xii\,a^3}+\Omega_0
+\Omega_6\,\frac{6\xii\,\Omega_0-10\vareps}{\omega(\omega\pm\Omega_3)}+{\cal O}(a^3),~~
\Psi_{\pm}=\frac{2}{\Omega_3\pm\omega}+{\cal O}(a^3),
\eea
where $\h,\h_{-}$ are stable and $\h_{+}$ is unstable. 

Let us finally assume that $\underline{\Omega_3=\Omega_4=0}$. Then there is only one solution, 
\be
\h=\frac{\Omega_2}{3a^2}+\frac{\eeta}{3}+\frac{8\Omega_2^3}{81\Omega_6}
+\left(
\frac{4\Omega_2^2(\Omega_0+\eeta)} {27\Omega_6}-\frac{32\,\Omega_2^5 }{729\, \Omega_6^2}
\right)a^2+{\cal O}(a^4),~~\Psi=\frac{3}{2\Omega_2 a}+{\cal O}(a),
\ee 
and it is stable. For $\Omega_2=0$ it remains stable and reduces to 
\be                     \label{fin}
\h=\frac{\eeta}{3}+\frac{4\eeta^2(3\Omega_0-\eeta)}{81\Omega_6}\,a^6+{\cal O}\left(a^{12}\right),~~~~
~\Psi=\frac{3}{2\eeta a^3}+{\cal O}\left(a^{3}\right).
\ee

Let us summarize  the above results in the case where 
$K=\Omega_2=0$. Assuming generic values of the other parameters, 
the stable near the singularity solution is 
\be                  \label{zero1}
\h=\frac{\eeta}{3}+\ldots,~~~~~~\Psi=\frac{3}{2\eeta a^3}+\ldots,
\ee
which is of the inflationary type. 
If $\Omega_4\neq 0$ then there is one more ghost-free solution 
\be                  \label{zero2}
\h=\frac{3\Omega_6}{\Omega_4 a^2}+\ldots,~~~~~~\Psi=-\frac{\Omega_4}{3\Omega_6 a}+\ldots,
\ee
hence $a(t)\sim t$; 
and if $\Omega_4=0$ but $\omega^2=\Omega_3^2-12\Omega_6>0$ 
this solution reduces to 
\be                   \label{zero3}
\h=\frac{\Omega_3-\omega}{2\,a^3}+\ldots,~~~\Psi=\frac{2}{\Omega_3-\omega}+\ldots,
\ee
hence $a(t)\sim t^{2/3}$. 
We note that the standard radiation-dominated solution \eqref{GR0} 
has ghost and is eliminated from consideration, while 
the ghost-free solutions \eqref{zero1}--\eqref{zero3} do not show a radiation-dominated phase
with $\h\sim 1/a^4$. Without this phase the model cannot correctly describe the primary 
nucleo-synthesis.

\subsection{Global solutions}

\begin{figure}[h]
\hbox to \linewidth{ \hss

	\resizebox{7cm}{6cm}
	{\includegraphics{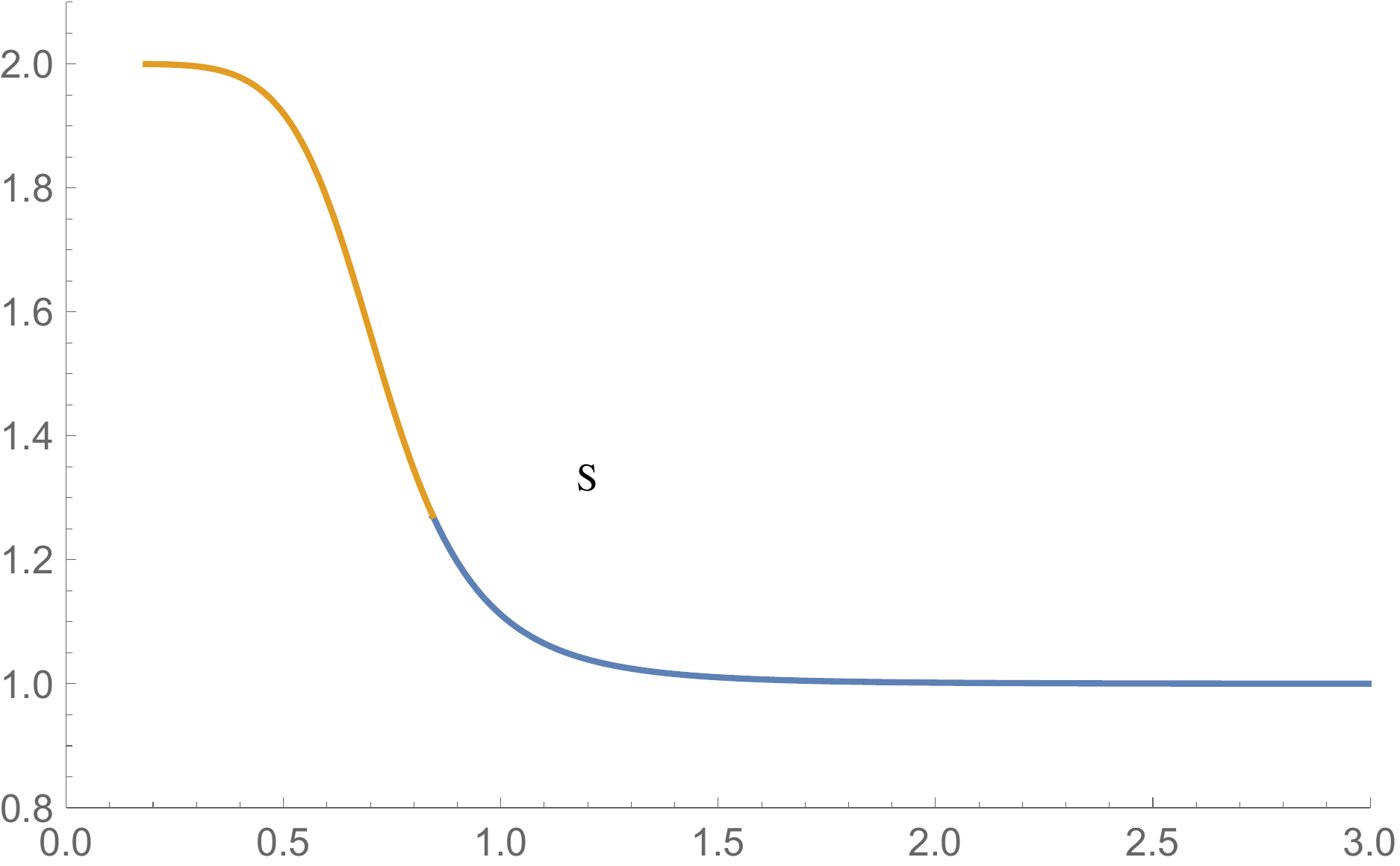}}
\hspace{5mm}
	\resizebox{7cm}{6cm}{\includegraphics{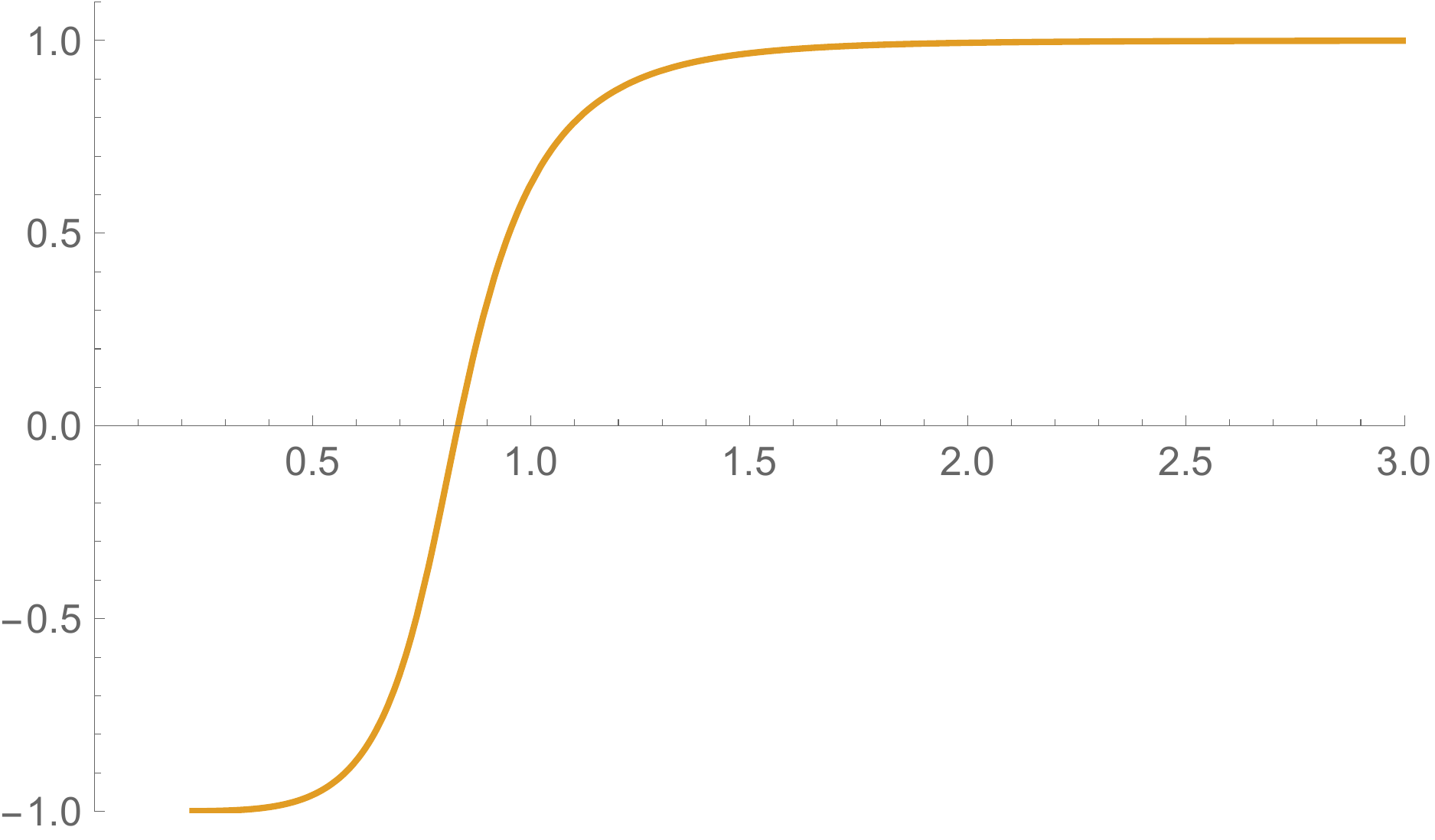}}
	
\hspace{1mm}
\hss}
\caption{
Solutions $\h(a)$ for 
$\Omega_0=\Omega_6=1$, $\Omega_2=0$, $\Omega_3=\Omega_4=0$
and for  $\vareps=6$ (left panel) or $\vareps=-3$ (right panel).
}
 \label{Fig1}
\end{figure}
We now know that the ghost-free solutions are 
described by 
Eqs.\eqref{zero1}--\eqref{zero3} near the singularity and by Eqs.\eqref{infty1},\eqref{infty2}
at late times. 
Having understood their asymptotic structure,  we can construct the solutions globally,
for example by using the Cardano formula. We shall start by describing  all possible solution types and later
select those which are ghost-free.

Let us assume at first that there is no matter, 
$\Omega_3=\Omega_4=0$. Then there is only one solution \eqref{fin} near the singularity. 
One can always rescale the system to set $\Omega_0=1$. It turns out that the 
qualitative behaviour of the solutions is insensitive to the value of $\Omega_6$, as long as $\Omega_6>0$,
hence one can set $\Omega_6=1$. 
At the same time, 
the value of $\eeta$ is important as it determines the number of 
solutions at infinity. If $\eeta>\Omega_0=1$ or $\eeta<0$ then there is just one solution at infinity, and hence 
only one global solution. If $\eeta>\Omega_0$ then this is the solution of the type S shown in Fig.\ref{Fig1}. 
It has  asymptotics
\be                  \label{s1}
\frac{\eeta}{3}\leftarrow \h \to \Omega_0\,,
\ee
hence the universe inflates with constant but different Hubble rates at early times and at late times. 
The function ${\cal G}$ is everywhere positive, therefore this solution is stable. 

If $\eeta<0$ then the  solution is described by the curve of type 
shown in the right panel of Fig.\ref{Fig1}. Such a solution makes sense only in the region 
$a>a_{\rm min}$ where $\h=(H/H_0)^2>0$.
Since $H=\pm H_0\sqrt{\h}$ can have both signs in this region, 
the solution describes a  bounce -- a universe contracting with $H=-H_0\sqrt{\h}$
up to a minimal size $a_{\rm min}$ and then expanding with  $H=+H_0\sqrt{\h}$. 
(This behaviour is similar to that found in the scalar-tensor gravity
with a negative scalar field potential \cite{Boisseau:2015hqa}.)
However, such bounce solutions are unstable, since ghost is present for $\eeta<0$.

Let us now choose a value $\eeta\in(0,\Omega_0)$.  
\begin{figure}[h]
\hbox to \linewidth{ \hss

	\resizebox{7cm}{6cm}
	{\includegraphics{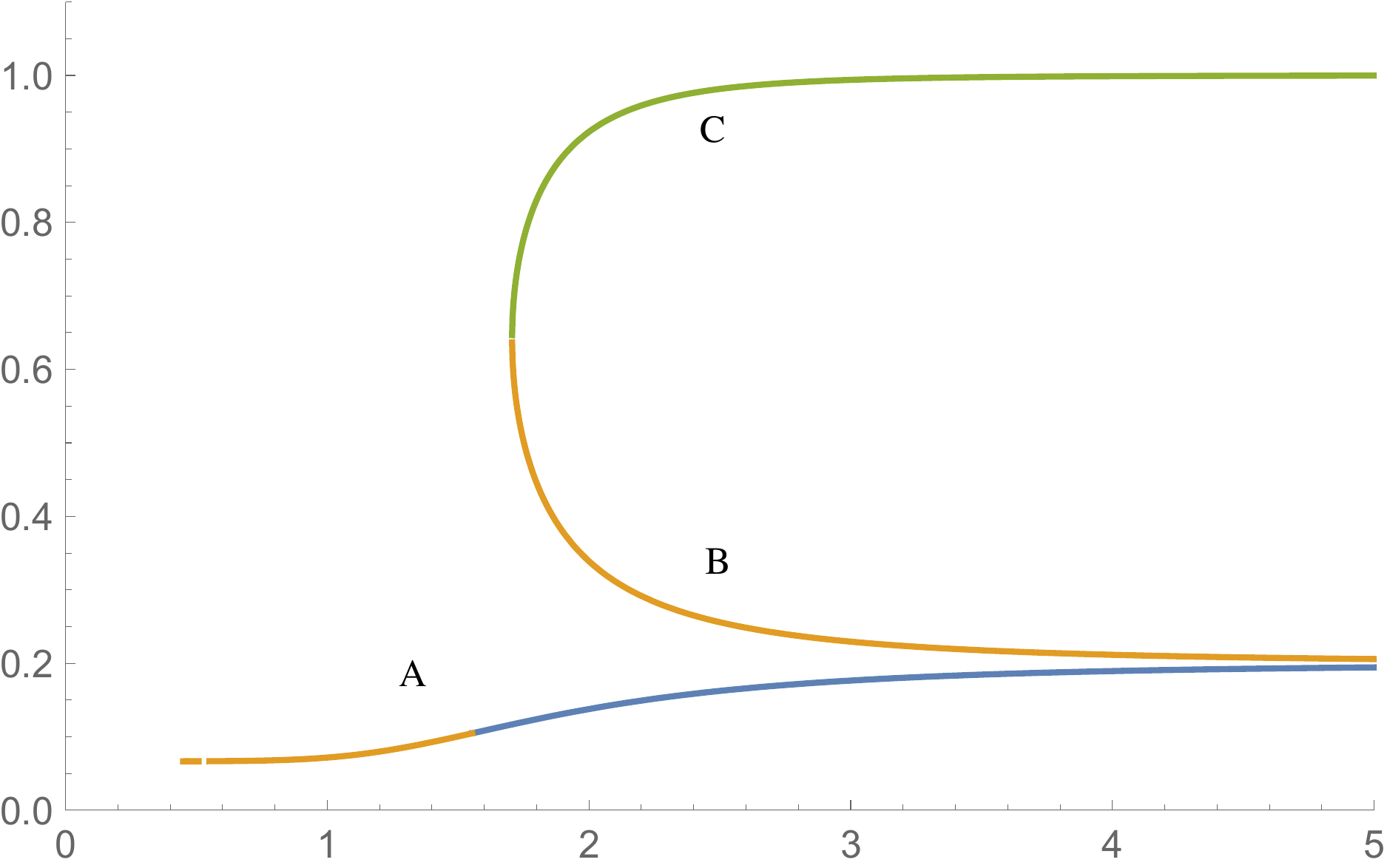}}
\hspace{5mm}
	\resizebox{7cm}{6cm}{\includegraphics{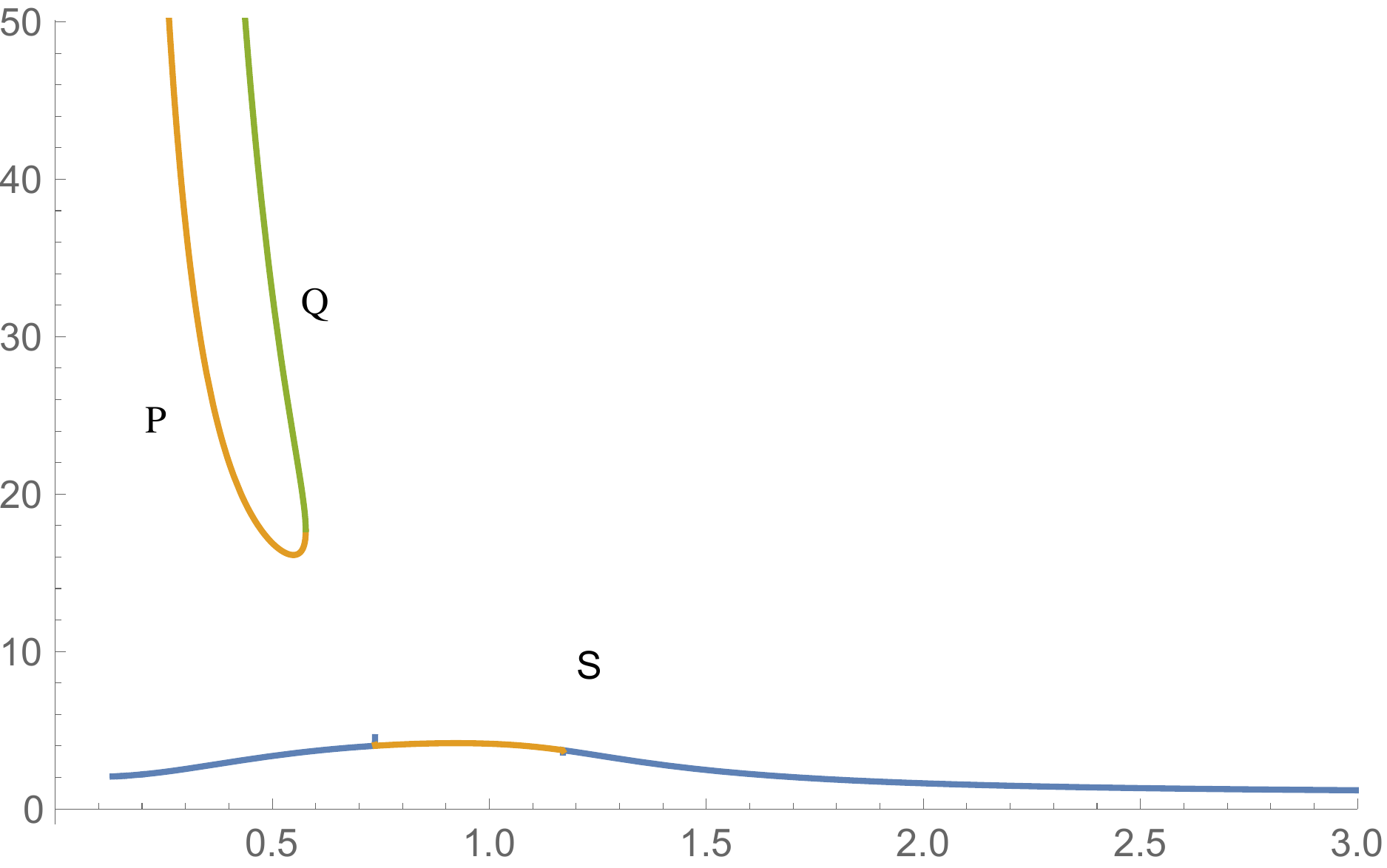}}
	
\hspace{1mm}
\hss}
\caption{
Solutions $\h(a)$ for $\Omega_0=\Omega_6=1$, $\Omega_2=0$, $\Omega_3=\Omega_4=0$, $\vareps=0.2$ (left)
and for $\Omega_0=\Omega_6=1$, $\Omega_2=0$, $\Omega_3=5$, $\Omega_4=0$, $\vareps=6$ (right). 
}
 \label{Fig2}
\end{figure}
Then there are three local solutions at  large $a$ and only one  at small $a$, hence there is only one
solution that continues from the large $a$ all the way down to small $a$.  This corresponds to the solution A
 in Fig.\ref{Fig2}; it has  asymptotics 
\be               \label{s2}
\frac{\eeta}{3}\leftarrow \h \to \eeta,
\ee
and it is stable. The two other solutions, $B$ and $C$ in  Fig.\ref{Fig2}, exist only near infinity and cannot extend
down to the singularity, hence they merge each other at some finite value $a=a_\ast$ where $\h(a_\ast)=\h_\ast$. 
One has near this point $a-a_\ast\sim (\h-\h_\ast)^2$ and hence 
\be
a(t)=a_\ast+a_\ast\sqrt{\h_\ast}|t-t_\ast|+{\cal O}\left( (t-t_\ast)^{3/2}\right).
\ee
The geometry is singular  at the moment $t=t_\ast$  when the time ``emerges" 
and it is not possible to regularly continue 
the spacetime to the $a<a_\ast$ region. (A similar situation occurs in the $F(R)$ gravity if $F^{\prime\prime}(R)=0$ 
at some $R=R_\ast$  \cite{Appleby:2009uf}.)
Such solutions are unlikely 
to be physically interesting,  in addition the solution $C$ is unstable. 
Summarizing, only the solutions  S in Fig.\ref{Fig1} and A in Fig.\ref{Fig2}  are stable.

Let us now add the matter by setting $\Omega_4\neq 0$ and/or $\Omega_3\neq 0$. This does not 
affect much solutions at large $a$, but this creates new solutions at small $a$. Let us again 
consider the $\eeta>\Omega_0$ case. Then there is only one solution at infinity 
but there are three of them near the singularity. Therefore, only one solution can extend to the whole interval of $a$,
this is the solution $S$ shown in the right panel of  Fig.\ref{Fig2}. This solution is similar 
to the S in  Fig.\ref{Fig1}, since it is also stable and has the same boundary conditions \eqref{s1},  while
taking the matter into account only produces  some deformations of the solution in the intermediate region. 
At the same time, the matter gives rise to two more solution near the singularity
-- solutions $P$ and $Q$ in the right panel 
of  Fig.\ref{Fig2} -- they 
cannot extend to large $a$ and hence merge each other at some point (``the end of time"). 
It is again unlikely that such solutions could by physically interesting, 
because one of then is unstable. Therefore, including the matter does not 
bring anything new at this point.

\begin{figure}[h]
\hbox to \linewidth{ \hss

	\resizebox{7cm}{6cm}
	{\includegraphics{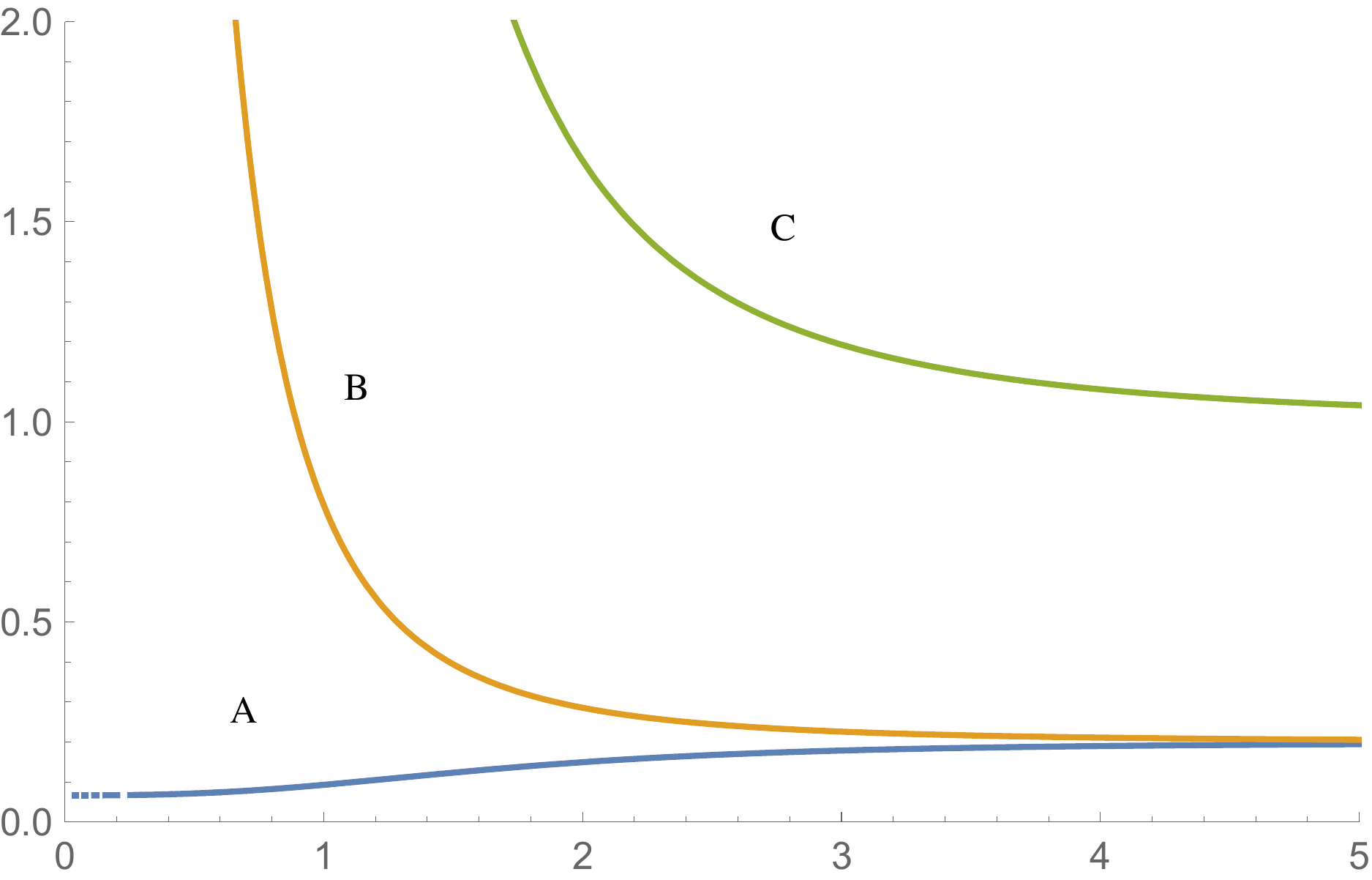}}
\hspace{5mm}
	\resizebox{7cm}{6cm}{\includegraphics{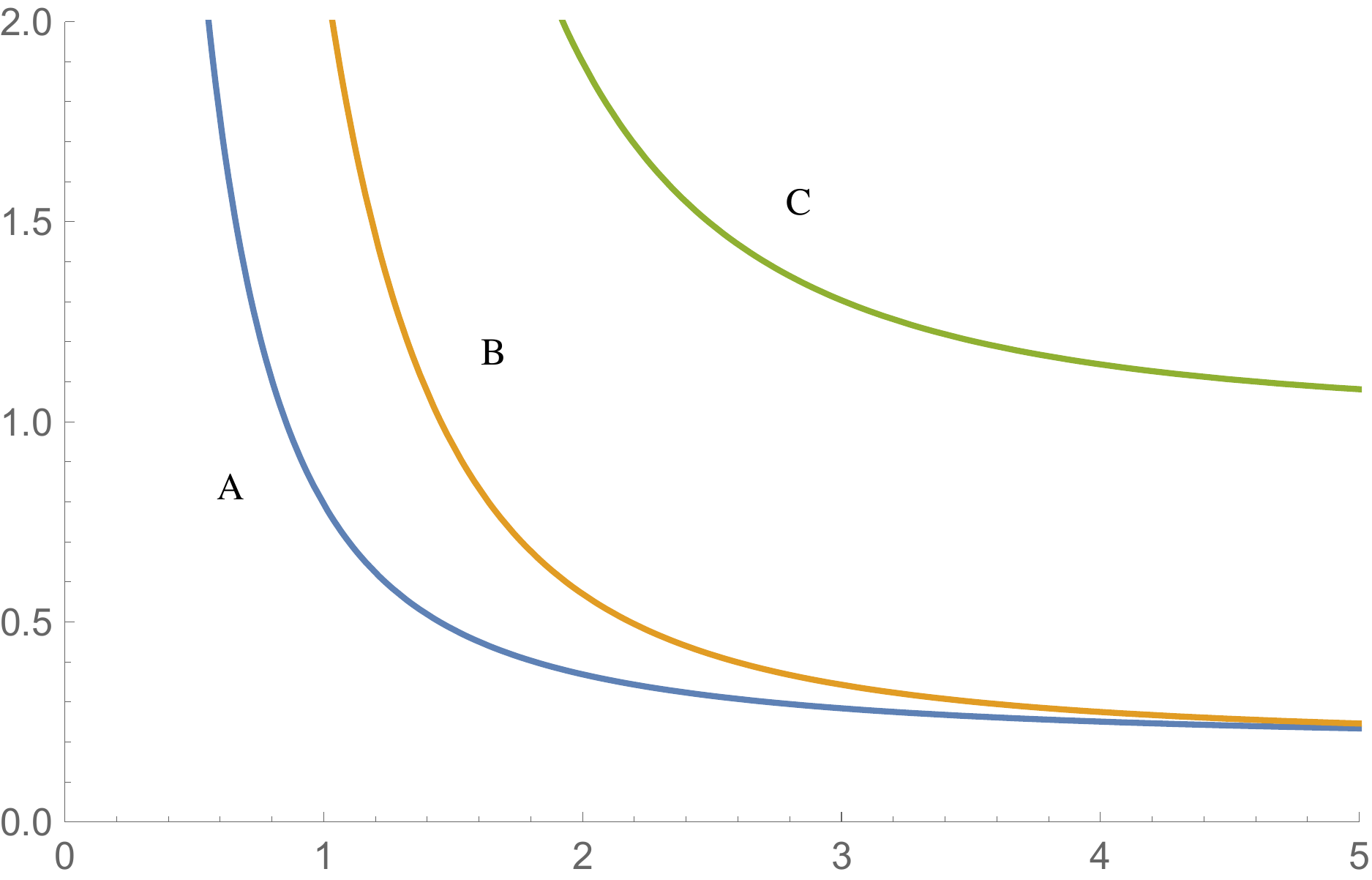}}
	
\hspace{1mm}
\hss}
\caption{
Solutions $\h(a)$ for $\Omega_0=\Omega_6=1$,  $\Omega_3=5$, $\Omega_4=0$, $\vareps=0.2$. 
One has $\Omega_2=0$ (left)
and $\Omega_2=1$ (right). 
}
 \label{Fig3}
\end{figure}
However, the matter  can be essential if $\eeta\in(0,\Omega_0)$, since in this case the
number of solutions at small $a$  matches that at large $a$, hence there can be three different solutions 
extending to the whole interval of $a$. This is shown in  Fig.\ref{Fig3}, where the solution A is essentially
the same as before, in  Fig.\ref{Fig2}. However, the solutions $B$ and $C$ no longer merge each other but extend 
all the way down to the singularity at $a=0$ where they meet the corresponding local solutions. 
The solution $C$ is unstable but the  B is stable. Therefore, in addition 
to the solutions of types S and A with the asymptotics \eqref{s1} and \eqref{s2}, there is a third stable global solution 
with the following behaviour, 
\be                                  \label{s3}
\frac{3\Omega_6}{\Omega_4 a^2}\leftarrow \h\to \eeta~~~\mbox{or} ~~~~~~
\frac{\Omega_3-\sqrt{\Omega_3^2-12\Omega_6}}{2 a^3}\leftarrow \h\to \eeta~~~\mbox{if}~~\Omega_4=0.
\ee 
This exhaust all ghost-free solutions. 
All the above arguments applies also for $\Omega_2\neq 0$, for example the 
three global solutions  for $\Omega_2>0$
are shown in the right panel of Fig.\ref{Fig3}.

Let us finally note that,
since one has at present $\h=a=1$, one should have 
\be                 \label{norm}
1=\Omega_0+{\Omega_2}+{\Omega_3}+{\Omega_4}
+\Omega_6\,\frac{\vareps-3+{\Omega_2}  }{\left[\vareps-1+{\Omega_2} \right]^2  }. 
\ee
However, 
since there can be several solutions $\h(a)$, one cannot impose 
this condition beforehand. Indeed, the condition requires 
the $\h(a)$ curve to pass through the physical point $(1,1)$, but one does not know in advance 
which of the several curves 
is physical. Therefore, 
one has to first choose a particular solution $\h(a)$ obtained for some parameter values $\Omega_k,\vareps$,
then choose a point $(a_\ast,\h_\ast)$ on this curve, and then declare this point to be physical. 
Since one has 
\be                          \label{HHH1}
\h_\ast=\Omega_0+\frac{\Omega_2}{a_\ast^2}+\frac{\Omega_3}{a_\ast^3}+\frac{\Omega_4}{a_\ast^4}
+\frac{\Omega_6 \left[\vareps-3\h_\ast
+\frac{\Omega_2}{a_\ast^2}  \right]}{a^6 \left[\vareps-\h_\ast+\frac{\Omega_2}{a_\ast^2}  \right] ^2},
\ee
this can be rewritten as the normalization condition 
\be                 \label{norm1}
1=\Omega^\ast_0+{\Omega^\ast_2}+{\Omega^\ast_3}+{\Omega^\ast_4}
+\Omega^\ast_6\,\frac{\vareps^\ast-3+{\Omega^\ast_2}  }{\left[\vareps^\ast-1+{\Omega^\ast_2} \right]^2  },
\ee
with the modified parameter values 
\be
\Omega^\ast_0=\frac{\Omega_0}{\h_\ast},~~~
\Omega^\ast_2=\frac{\Omega_2}{\h_\ast a_\ast^2},~~~
\Omega^\ast_3=\frac{\Omega_3}{\h_\ast a_\ast^3},~~~
\Omega^\ast_4=\frac{\Omega_4}{\h_\ast a_\ast^4},~~~
\Omega^\ast_6=\frac{\Omega_6}{\h^2_\ast a_\ast^4},~~~
\vareps^\ast=\frac{\vareps}{\h_\ast}.
\ee
As a result, imposing the normalization condition is achieved by rescaling the parameters.

\section{Stability of the solutions}

Summarizing what was said above, the F5 theory \eqref{Fab5}  admits various 
cosmological solutions, but ghost-free solutions exist only 
if $\kkappa\geq 0$ and $\varepsilon\geq 0$.  The no-ghost conditions eliminate
many solutions, as for example the bounces or the ``emerging time" solutions. 
Setting $\varepsilon=0$ gives a subset of the F4 theory where 
flat space is a ghost-free solution, but it may develop ghost within the full F4, 
as shown in Appendix B. 
Expanding cosmologies 
are obtained for $\varepsilon>0$, and some of them still 
show ghost, but there are three different types 
of ghost-free solutions corresponding to  the curves S, A, B in Figs.\eqref{Fig1}--\eqref{Fig3}. 

The S solution exists for $\eeta>\Omega_0$ and is sourced  by the scalar field, 
but it may also contain the matter. It
describes a universe with the standard late time dynamic dominated by the $\Lambda$-term, 
radiation and dust. At early times the matter effects are totally screened and the universe expands with a
constant Hubble rate determined by $\varepsilon/\kkappa$. 
Since it contains two independent parameters $\eeta$ and $\Omega_0\sim\Lambda$ in the asymptotics, 
this solution can have an hierarchy between the Hubble scales  at the early and  late times. 
However, at late times it is not screening and dominated by $\Lambda$, 
thus invoking again the cosmological constant problem. 

The solutions A and B exist for $0<\eeta<\Omega_0$. The solution A is sourced by the scalar field,
with or without the matter, while the solution B exists only when the matter is present. 
They both show the screening 
because their late time behaviour is controlled 
by $\eeta\sim\varepsilon/\kkappa$ and not by $\Lambda$. Therefore, they could in principle
describe the late time acceleration while circumventing  the cosmological constant problem,
and one might probably find arguments justifying that $\varepsilon/\kkappa$ should be small. 
At the same time, these solutions cannot  describe  the early inflationary phase.
Indeed, the  near singularity behaviour  \eqref{s3} of the solution B does not correspond to inflation, 
while the solution A does show an inflationary phase, but with essentially the same Hubble rate 
as at late times, hence there is no hierarchy between the two Hubble scales. 
In addition, as was mentioned above, none of the ghost-free solutions has a 
radiation-dominated phase needed for the nucleo-synthesis. 
As we shall now see,
there are other problems near the singularity.  

The solutions S, A, B do not have the ghost instability. 
However, this does not mean they are stable, since they may have other instabilities. 
Let us therefore study their stability.  
The linear equations for generic perturbations of spatially flat
homogeneous and isotropic backgrounds are 
derived in Appendix \ref{ppert} (the $K=0$ case is the most important). 
These equations describe the scalar and tensor perturbations, but 
we shall illustrate the procedure by discussing only the tensor sector, 
 since the equations are simpler in this case. Each of the two  
 graviton polarizations is described by the same equation 
\eqref{tens}, 
 \bea
&&(\Omega_6\Psi^2+1)\,\ddot{w}+                      \label{tens0}
\left(2\Omega_6 \Psi\dot{\Psi}+3(\Omega_6\Psi^2+1)h\right)\,\dot{w} \\
&&-\left(
2(\Omega_6\Psi^2+1)(2\dot{h}+3h^2)
+2\Omega_6(3\eeta\Psi^2+4h\Psi\dot{\Psi})-6\Omega_0
+\frac{P^2}{a^2}\,(\Omega_6 \Psi^2-1)
\right)w=0,  \nonumber
\eea
where the derivatives are taken with respect to the dimensionless time $\tau=H_0\, t$. 
The equations in the scalar sector (Eqs.\eqref{scal1}--\eqref{scal3})  
are  more complicated but their solutions
are qualitatively similar. 

The coefficients in \eqref{tens0} depend on the Hubble parameter $h$ 
and scalar  $\Phi$, which are determined by the 
background solutions. The solutions will be stable if their 
perturbations are bounded. Now, since their equations
are linear, the perturbations can become unbounded only asymptotically, 
either in the past, when $a\to 0$, or in the future, when $a\to\infty$. 
It is therefore sufficient to study the perturbations only in the 
small time limit when the background solutions are described by Eqs.\eqref{zero1}--\eqref{zero3}, 
and also in the late time limit when the backgrounds are described by  \eqref{infty1} or by  \eqref{infty2}.

Let us first check if solutions of \eqref{tens0}  are bounded 
for $a \to\infty$. At late times one can neglect the $P^2/a^2$ term, 
while $h,\Psi$ are then given either by \eqref{infty1} or by \eqref{infty2}. 
Consider first the GR branch \eqref{infty1}. One has in this case $\Psi\sim 1/a^3$, hence 
for $a\to\infty$  one can set $\Psi=0$, while
$h=\sqrt{\Omega_0}$. As a result, Eq.\eqref{tens0} reduces to 
$
\ddot{w}+3h\dot{w}=0,
$
whose solution 
\be
w=C_1+C_2\, e^{-3h\tau }\,
\ee
is bounded as $\tau\to \infty$. Therefore, all tensor modes are bounded at late times. 
Considering similarly the 
scalar perturbation modes $w,u,\phi$ described by Eqs.\eqref{scal1}--\eqref{scal3}  
one finds 
\be
w=C_1,~~~~~u=C_2,~~~~~~\phi=C_3+C_4 e^{-3h\tau},
\ee
and hence  $w,u$ and $\dot{\phi}/\Psi$ are bounded, too. The conclusion is that the 
GR branch \eqref{infty1} is dynamically stable.

For  the screening branch \eqref{infty2} one has 
\be
h=\sqrt{\eeta},~~~~~~~\Psi=\pm \sqrt{\frac{\Omega_0-\eeta}{2\eeta\Omega_6}}\,,
\ee
injecting which to Eq.\eqref{tens0} reduces the equation again to  $\ddot{w}+3h\dot{w}=0$,
 hence the solution is again 
 $
w=C_1+C_2 e^{-3h\tau }\,.
$
Considering the scalar modes $w,u,\phi$, 
one finds that they are bounded for $\tau\to\infty$ as well. 
Therefore, the screening branch is stable, too. 

Let us now see if the solutions are stable in the past, when $a\to 0$. 
The corresponding background solutions can be of three different 
types described by \eqref{zero1}--\eqref{zero3}. 
For small $a$ one cannot neglect  the $P^2/a^2$ term anymore, but 
let us first analyse the homogeneous modes with $P=0$. 
For the screening  branch described by \eqref{zero1} one has 
\be                  \label{pp1}
h=\sqrt{\eeta/3},~~~~~\Psi=\frac{3}{\eeta a^3},
\ee
inserting which to Eq.\eqref{tens0} gives
\be                 \label{pp2}
\frac{\Omega_6}{h^4 a^6}\left(\ddot{w}-3\dot{w}\right)
+\ddot{w}+3h\dot{w}+6(\Omega_0-h^2)w=0. 
\ee
When $a$ is small, the term proportional to $1/a^6$ is dominant and hence 
the equation reduces to 
 $
\ddot{w}-3h\dot{w}=0
$
whose solution 
$
w=C_1+C_2 e^{+3h\tau } 
$
is bounded for  $\tau\to -\infty$. The scalar sector amplitudes $w,u,\phi$ are bounded as well. 
Therefore, this branch is stable with respect to homogeneous perturbations. 

For the second branch described by \eqref{zero2} one has
\be
a(\tau )=\sqrt{\frac{3\Omega_6}{\Omega_4}}\,\tau ,~~~~~~~~\Psi=-\frac{\Omega_4}{3\Omega_6 a}\,,
\ee
inserting which to Eq.\eqref{tens0} and keeping only the leading at small $a$ 
terms gives 
\be
\ddot{w}+\frac{1}{t}\,\dot{w}+\frac{6}{t^2}\,w=0.
\ee
The solution is
$
w=C_1\,\cos(\sqrt{6}\ln(\tau)+C_2)
$
and this is bounded as $\tau\to 0$, however its derivatives grow without bounds
and hence the curvature blows up. In addition, solving for the scalar modes $w,u,\phi$ 
one finds that $\phi$ contains a piece proportional to $1/\tau^3$, hence $\dot{\phi}/\Psi$
is unbounded. Since the perturbations grow, 
this solution branch  is unstable. 

For  the third branch \eqref{zero3} one has $a(\tau)\sim \tau^{2/3}$ and 
$\Psi={\cal O}(1)$,  in which case Eq.\eqref{tens0} yields $w\sim 1/\tau$, 
hence this branch is unstable, too. 

Summarizing, so far only the 
screening branch \eqref{zero1} passes the stability check.
  This branch is actually the most 
interesting, however, its stability is not yet established, since 
there remains to consider the inhomogeneous perturbations with $P\neq 0$. 
Let us therefore return to Eq.\eqref{pp1}
and insert it to Eq.\eqref{tens0}. The result is
\be                 \label{pp22}
\frac{\Omega_6}{h^4 a^6}\left(\ddot{w}-3\dot{w}-\frac{P^2}{a^2}\,w \right)
+\ddot{w}+3h\dot{w}+6(\Omega_0-h^2)w+\frac{P^2}{a^2}\,w=0,
\ee
and it is clear that the terms proportional to $\Omega_6$ are dominant 
when $a\to 0$, hence the equation reduces to 
\be
\ddot{w}-3\dot{w}-\frac{P^2}{a^2(\tau)}\,w=0~~~~~\mbox{with}~~~~~a(\tau)=e^{h\tau}. 
\ee
The solution of this equation
\be
w=C_1\, a^2(\tau)[P-ha(\tau)]\exp\left(\frac{P}{ha(\tau)}\right)
+C_2\, a^2(\tau)[P+ha(\tau)]\exp\left(-\frac{P}{ha(\tau)}\right)
\ee
diverges as $a(\tau)\to 0$, and the divergence is very strong -- it is proportional to 
the exponent of exponent of $\tau$. This effect is produced 
 by terms proportional to $\Omega_6$, hence by the background
scalar. Therefore, the screening solution \eqref{pp1} is 
unstable as well.

As a result, we conclude that all isotropic ghost-free solutions in the theory are unstable in the vicinity
of the initial spacetime singularity. We know that  in GR the isotropic
cosmologies are also unstable near the singularity 
(hence one should study more general
anisotropic solutions),  but in our case the instability is stronger -- it is exponential and not 
power law as in GR.  
Therefore, the F5 theory does not have viable isotropic solutions describing
the whole of the cosmological history. This conclusion is supported by the 
previous observation that the model does not have a radiation-dominated phase
and hence cannot correctly describe the primary nucleo-synthesis. 
On the other hand, one cannot perhaps expect a particular
field theory model to be able to describe everything, whereas at late times it
admits stable solutions with an accelerating phase. For 
$0<\eeta<\Omega_0$ these  solutions show the screening, since their Hubble 
 parameter is determined not by
the conventional $\Lambda$-term but by $\eeta\sim \varepsilon/\kkappa$, 
which circumvents the cosmological constant problem.  Hence the model 
fulfills the necessary for it condition -- to provide an explanation for the current cosmic acceleration. 

Therefore, the main outcome of our analysis is the conclusion that the Horndeski theory 
may indeed offer interesting for cosmology features. Although the model we considered
is not totally satisfactory, one may hope that more realistic models can be obtained by adjusting 
the coefficient functions $G_k(X,\Phi)$ in the Horndeski Lagrangian \eqref{horn}. 
Unfortunately, to decide whether or not a given model is realistic  always requires to carry out a tedious 
analysis, similar to the one presented above.

 \acknowledgments

We thank Gary Gibbons for discussions and a reading of the manuscript.  
MSV was partly supported by the Russian Government Program of Competitive Growth 
of the Kazan Federal University. 
AAS and SVS were supported by the RSF grant 16-12-10401.
SVS kindly appreciates the hospitality of the LMPT at the University of Tours. 
 
\appendix
\setcounter{section}{0}
\setcounter{equation}{0}
\setcounter{subsection}{0}
\section{The no-ghost conditions \label{G}}

\renewcommand{\theequation}{\Alph{section}.\arabic{equation}}

These conditions guarantee that the kinetic energy is positive. At the perturbative level, 
this means that the kinetic part of the second variation of the Lagrangian  should be described by 
a positive-definite quadratic form. 
The second variation of the Lagrangian can be obtained by perturbing 
the background  configuration,
\be
g_{\mu\nu}\to g_{\mu\nu}+\delta g_{\mu\nu},~~~~~\Phi\to \Phi+\delta \Phi, 
\ee 
computing the linearized field equations for the perturbations $\delta g_{\mu\nu}$ and $\delta\Phi$, 
\bea                     \label{Eeqs}
\delta E_{\mu\nu}&\equiv& \delta(\mmu\, G_{\mu\nu}+\Lambda g_{\mu\nu}-\kkappa \,{\cal T}_{\mu\nu}-\varepsilon\, T^{(\Phi)}_{\mu\nu}-T^{\rm (m)}_{\mu\nu}), \nonumber \\
\delta E_\Phi&\equiv& 
\delta(\nabla_\mu((\kkappa G^{\mu\nu}+\varepsilon g^{\mu\nu})\nabla_\nu\Phi)), 
\eea
and  using  this to compute 
\be                                    \label{ddS}
\delta^2 S=\frac12\int ( \delta E_{\mu\nu}\,\delta g^{\mu\nu}+\delta E_\Phi\, \delta \Phi)\sqrt{-g}\,d^4 x\equiv 
\frac12\int \delta^2 L\, d^4x. 
\ee
The resulting $\delta^2 L$ splits into three independent parts corresponding to contributions
of the scalar, vector, and tensor modes, and the positivity of the kinetic terms imposes an independent 
condition in each sector. 

Before considering generic perturbations, it is instructive to see that 
there is a simple way to obtain the correct answer 
by considering only anisotropic perturbations. 
Let us assume  the spacetime metric to be 
\be                        \label{ma}
ds^2=-N^2\,dt^2+a_1^2\,dx^2+a_2^2\,dx_2^2+a_3^2\,dx_3^2\,,
\ee
where $N,a_k$ are functions of $t$, while  $\Phi=\Phi(t)$. This can be viewed 
as an anisotropic deformation of the isotropic metric with $K=0$, 
and  for the generic $a_k$ such a deformation contains contributions from both 
scalar and tensor perturbation sectors. 
Dropping a total derivative
  and neglecting the matter contribution, 
the Lagrangian \eqref{Fab5}  is 
\be                    \label{LL}
L=-\left(\frac{2\mmu}{N}
+\frac{\kkappa\, \dot{\Phi}^2}{N^3}\right)Q
+\left(\frac{\varepsilon \, \dot{\Phi}^2}{N}-2N\Lambda\right)\,a_1\, a_2\, a_3\,,
\ee
with 
\be                          \label{Q}
Q=a_1\,\dot{a}_2\,\dot{a}_3
+a_2\,\dot{a}_1\,\dot{a}_3
+a_3\,\dot{a}_1\,\dot{a}_2\,.
\ee
Varying $L$ with respect to the lapse $N$ gives 
the first order constraint, 
\be                    \label{CC}
{\cal C}=\left(\frac{2\mmu}{N^2}
+\frac{3\kkappa\, \dot{\Phi}^2}{N^4}\right)Q
-\left(\frac{\varepsilon \, \dot{\Phi}^2}{N^2}+2\Lambda\right)\,a_1\, a_2\, a_3=0, 
\ee
whereas varying with respect to $a_k$ and $\Phi$ gives the second order equations.
Let us set $N=1$ and assume the configuration to be almost isotropic, 
$$
a_k=a+\delta a_k,~~~~~
\Phi=\int \psi \,dt+\delta\Phi.
$$
Then $Q=3a\dot{a}^2+\delta Q+\ldots \equiv Q_0+\delta Q+\ldots$ with 
\be
\delta Q=2a\dot{a}(\delta\dot{a}_1+\delta\dot{a}_2+\delta\dot{a}_3)
+\dot{a}^2(\delta a_1+\delta a_2+\delta a_3),
\ee
and
$
{\cal C}=(2\mmu+3\kkappa\psi^2)3a\dot{a}^2
-(\varepsilon\psi^2+2\Lambda)a^3+\delta{\cal C}+\ldots \equiv 
{\cal C}_0+\delta{\cal C}+\ldots
$
with
\be                                    \label{CCC}
\delta{\cal C}=(2\mmu+3\kkappa\psi^2)\delta Q+
2(9\kkappa a\dot{a}^2-\varepsilon a^3)\psi\delta\dot{\Phi}
-(\varepsilon\dot{\Phi}^2+2\Lambda)a^2(\delta a_1+\delta a_2+\delta a_3),
\ee
while 
the Lagrangian \eqref{LL} expands as 
$
L=L_0+\frac12\,\delta^2L+\ldots
$
 where $\delta^2 L$ is a quadratic in $\delta\dot{\Phi}$, 
 $\delta\dot{a}_k$, $\delta a_k$ form. The first order condition $\delta{\cal C}=0$ 
 can be used to express $\delta\dot{\Phi}$ in terms of the other variables, 
\be                    \label{W}
\delta\dot{\Phi}={\cal W}\,(\delta\dot{a}_1+\delta\dot{a}_2+\delta\dot{a}_3)+\ldots\,,
\ee
where the dots denote terms without derivatives and 
\be                   \label{WW}
{\cal W}=\frac{(2\mmu+3\kkappa \psi^2)H  }{(\varepsilon-9\kkappa H^2  )a\psi}\,.
\ee
Inserting \eqref{W} to $\delta^2 L$  gives 
a quadratic from containing only  $\delta\dot{a}_k$ and $\delta a_k$, 
\be                             \label{dLL}
\delta^2L=A(\delta\dot{a}_1+\delta\dot{a}_2+\delta\dot{a}_3)^2
+2B(\delta\dot{a}_1\delta\dot{a}_2+\delta\dot{a}_1\delta\dot{a}_3+\delta\dot{a}_2\delta\dot{a}_3)+\ldots\,,
\ee
where 
\be
A=(\varepsilon-3\kkappa H^2) a^3 {\cal W}^2-4\kkappa a^2 H {\cal W},~~~~~~B=-\frac12\,\,(2\mmu+\kkappa\,\psi^2)\,.
\ee
The form \eqref{dLL} has eigenvalues 
\be
\lambda_1=3A+2B,~~~~~~\lambda_2=\lambda_3=-B,
\ee
which should  be non-negative. Therefore, using the above values of $A,B,{\cal W}$
and requiring the kinetic term to be positive, 
one arrives at the no-ghost conditions
\be                         \label{g-s}
\lambda_1=\frac{a\,(18\kkappa H^2\psi^2+6\mmu H^2-\varepsilon\, \psi^2) 
(9\kkappa^2 H^2\psi^2-6\kkappa\mmu H^2+\varepsilon \,(\kkappa \psi^2 +2\mmu))
 }{\psi^2 \,(9\kkappa H^2-\varepsilon)^2  }>0
\ee
and 
\be                     \label{g-t}
\lambda_2=\lambda_3=\frac{a}{2}\,(2\mmu+\kkappa \psi^2)>0\,. 
\ee
It turns out that these  are, respectively,   the same 
conditions as those for the generic  scalar and tensor perturbations.

Let us now see what happens if $K\neq 0$. One notes first that 
the scalar sector condition \eqref{g-s} can  be obtained in a simpler way. 
Indeed, restricting to only isotropic perturbations,
$
\delta a_1=\delta a_2=\delta a_3=\delta a,
$
\eqref{dLL} reduces to 
\be
\delta^2 L=3(3A+2B)\delta\dot{a}^2=3\lambda_1 \delta\dot{a}^2+\ldots, 
\ee
hence $\lambda_1$ can be computed by perturbing only $\Phi$ and the scale factor $a$
 of the isotropic background.  This can be easily done for 
non-zero values of the spatial curvature $K$. 
Let us consider the FLRW metric  for an arbitrary $K$, 
\be                            \label{FLRW11}
ds^2=-N^2(t)\,dt^2+a^2(t)\,\left[
\frac{dr^2}{1-Kr^2}+r^2 (d\vartheta^2+\sin^2\vartheta d\varphi^2)
\right].
\ee
 Inserting this to \eqref{Fab5} 
yields the reduced Lagrangian 
\bea
L&=&-\frac{3\kkappa a }{N^3}\, \dot{\Phi}^2 (\dot{a}^2+KN^2)
+\frac{6\mmu a}{N}\left(
KN^2-\dot{a}^2\right)+\left(\frac{\varepsilon}{N}\,\dot{\Phi}^2-2\Lambda N \right)a^3
\eea
and the constraint ${\cal C}=\partial L/\partial N$. Setting $N=1$, perturbing the scale factor $a\to a+\delta a$
and the scalar field $\Phi\to \Phi+\delta\Phi$, expanding $L$ and ${\cal C}$ 
with respect to $\delta a,\delta \Phi$ and then using $\delta{\cal C}=0$ to express $\delta\dot{ \Phi}$ 
in terms of $\delta\dot{a}$, one obtains 
\be
\delta^2 L=3\lambda_1 \delta\dot{a}^2+\ldots
\ee
with 
\be                         \label{g-s1}
\lambda_1=\frac{a\,(18\kkappa H^2\psi^2+6\mmu H^2-\epsilon\, \psi^2) 
(9\kkappa^2 H^2\psi^2-6\kkappa\mmu H^2+\epsilon \,(\kkappa \psi^2 +2\mmu))
 }{\psi^2 \,(9\kkappa H^2-\epsilon)^2  },
\ee
where
\be                        \label{g-s2}
\epsilon=\varepsilon-\frac{3\kkappa K}{a^2}\,.
\ee 
This reduces to $\lambda_1$ in \eqref{g-s} if $K=0$. 
The ghost in the scalar sector  will be absent if $\lambda_1>0$.  

One may wonder why the no-ghost condition depends on $K$. 
Normally this is not the case, since changing $K$ changes the spatial curvature but
not the temporal derivatives of the metric, hence the metric kinetic 
term  does not depend on $K$. However, 
the kinetic term of the scalar field contains the non-minimal contribution 
 $G_{\mu\nu}\nabla^\mu\Phi\nabla^\nu\Phi$, and this depends on $K$, 
since the Einstein tensor does. 
 Therefore, the full kinetic term of the theory 
 depends on $K$, which is why  the scalar sector no-ghost condition is $K$-dependent. 
 At the same time, the no-ghost condition  in the tensor sector is expected to be the same for any $K$ 
 and to be given by \eqref{g-t}, because $\delta\Phi=0$ for the tensor (and vector) perturbations.

 The above conclusions are confirmed by analysing generic perturbations 
of the isotropic background. The corresponding equations for the perturbations 
are shown in Appendix \ref{ppert} below (for $K=0$). 
Inserting these equations to \eqref{ddS} and integrating by parts 
 to get rid of the second derivatives determines the second variation 
$\delta^2 L$, which splits into three independent parts  corresponding to contributions
of the scalar, vector and tensor modes. The kinetic part of $\delta^2 L$  in each sector is a 
manifestly positive definite quadratic form multiplied by a factor depending on the background amplitudes. 
The positivity of these factors for the tensor and vector modes requires 
of $\lambda_2$ defined by Eq.\eqref{g-t} is positive, while the positivity in the scalar sector requires that 
 $\lambda_1$ given by \eqref{g-s1},\eqref{g-s2} is positive. 

The conclusion is that ghosts will be absent if $\lambda_1$ and $\lambda_2$
defined by Eqs.\eqref{g-t},  \eqref{g-s1}, \eqref{g-s2}  
 are positive. Since $\psi$ is unbounded,  
the condition \eqref{g-t}, $2\mmu+\kkappa\psi^2>0$,
requires that $\kkappa>0$, which explains 
Eq.\eqref{gh1} in the main text. Eqs.\eqref{g-s1}, \eqref{g-s2}  
are equivalent to Eq.\eqref{gh2} in the main text. 

\appendix
\setcounter{section}{1}
\setcounter{equation}{0}
\setcounter{subsection}{0}
\section{Stability of flat space in the full F4 \label{GG}}

\renewcommand{\theequation}{\Alph{section}.\arabic{equation}}

For the sake of completeness, we derive here the no-ghost conditions for flat space in the full F4 theory. 
Choosing the metric  in the FLRW form \eqref{FLRW11} 
and $\Phi=\Phi(t)$,  the  Lagrangian \eqref{F40} becomes (up to a total derivative) 
\be
L_{\rm F4}=L_J+L_P+L_G+L_R-2\,\Lambda\, N a^3\,, 
\ee
with 
\bea
L_J&=&\frac{3a V_J}{N^3}\, \dot{\Phi}^2 (\dot{a}^2+KN^2)\,,\nonumber  \\
L_P&=&-\frac{3\dot{a}V_P}{N^5}\,\dot{\Phi}^3 \,(\dot{a}^2+KN^2)\,, \nonumber  \\
L_G&=&\frac{6a}{N}\left(
V_G\,(KN^2-\dot{a}^2)-V^\prime_G\,\dot{\Phi}\,a\dot{a}\right)\,, \nonumber  \\
L_R&=&-\frac{8\dot{a}\dot{\Phi}V^\prime_R }{N^3}(\dot{a}^2+3KN^2). 
\eea
Varying this with respect to $N,a,\Phi$ gives the constraint and the field equations. 
Whatever the functions $V_P(\Phi)$, $V_J(\Phi)$, $V_G(\Phi)$, $V_R(\Phi)$ are 
(unless $V_P=V_J=V^\prime _G=0$), 
the equations admit the flat space solution 
\be
N=1,~~~~~K=-1,~~~~~~a=t,
\ee
with the scalar field determined by 
\bea
V_P\,\dot{\Phi}^3-t\,V_J\,\dot{\Phi}^2+t^2\,V^\prime_G\,\dot{\Phi}=\frac{\Lambda}{3}\,t^3\,. 
\eea
The no-ghost conditions for this solution can be obtained by applying the described above procedure:
perturbing  the $K=-1$ metric to compute the kinetic term for the scalar modes, and  considering 
the anisotropic deformations  of the $K=0$ metric to describe the tensor modes. This shows that 
the scalar and tensor ghosts will be absent if the following two 
conditions hold,
\bea
9V_P\,\psi^3-5t V_J\,\psi^2+(2t^2V^\prime_{G}+8V^\prime_R)\psi+2tV_G>0,  \nonumber \\
3V_P\,\psi^3-t V_J\,\psi^2+8V^\prime_R\psi+2tV_G>0,
\eea
where $\psi=\dot{\Phi}$. Although these conditions are fulfilled if
$V_P=V_R=0$, $V_J=-\kkappa<0$, and $V_G=\mmu$,
as chosen in the main text, in general they  impose non-trivial conditions 
on the coefficient functions $V_J(\Phi),\ldots\,, V_R(\Phi)$. Therefore, flat space 
within the full F4  theory can be unstable
 and it will be ghost-free if only the coefficients $V_J,\ldots, V_R$ are properly chosen.

 \appendix
\setcounter{section}{2}
\setcounter{equation}{0}
\setcounter{subsection}{0}
\section{Equations for generic perturbations  \label{ppert} }

\renewcommand{\theequation}{\Alph{section}.\arabic{equation}}

Consider a homogeneous and isotropic background with $K=0$, 
\be
g^{(0)}_{\mu\nu}dx^\mu dx^\nu=-dt^2+\A^2(t)\, (d{\bf x})^2,~~~~~~~\Phi^{(0)}=\int \psi(t) dt\,.
\ee
Its  generic inhomogeneous and anisotropic perturbations can be expressed in the 
 synchronous gauge as
\be
g_{\mu\nu}=g^{(0)}_{\mu\nu}+\delta g_{\mu\nu},~~~~~~~\Phi=\int \psi(t)\, dt+\delta\Phi,
\ee
where 
\be
\delta g_{0\mu}=0,~~~~~\delta g_{ik}=2\A^2(t)\,h_{ik}(t)e^{i{\bf px}},~~~~~~~\delta\Phi=\bm \phi(t)e^{i{\bf px}}\,.
\ee
One can assume 
 the momentum vector to be oriented along the third axis, ${\bf p}=(0,0,p)$.  
The $h_{ik}$ tensor can be decomposed as 
\be
h_{ik}(t)=\sum_{m=1}^6 R_m(t) h^{(m)}_{ik}
\ee
where the basis matrices
\bea
h^{(1)}_{ik}&=&
\begin{pmatrix}
                             1 & 0& 0\\
                             0& 1 & 0\\
                             0& 0& 1
                             \end{pmatrix}\,,~~~
h^{(2)}_{ik}=
\begin{pmatrix}
                             1 & 0& 0\\
                             0& 1 & 0\\
                             0& 0& -2
                             \end{pmatrix}\,,~~~
h^{(3)}_{ik}=
\begin{pmatrix}
                             0 & 0& 1\\
                             0& 0 & 0\\
                             1& 0&  0
                             \end{pmatrix}\,,~~~\nonumber \\
h^{(4)}_{ik}&=&
\begin{pmatrix}
                             0 & 0& 0\\
                             0& 0 & 1\\
                             0& 1& 0
                             \end{pmatrix}\,,~~~
h^{(5)}_{ik}=
\begin{pmatrix}
                             1 & 0& 0\\
                             0& -1 & 0\\
                             0& 0& 0
                             \end{pmatrix}\,,~~~
h^{(6)}_{ik}=
\begin{pmatrix}
                             0 & 1& 0\\
                             1& 0 & 0\\
                             0& 0&  0
                             \end{pmatrix}\,.
\eea
Here $h_{ik}^{(1)}$ and $h_{ik}^{(2)}$ are diagonal and have 
a non-zero overlap with the ${\bf p}$ vector, they correspond to the scalar modes; 
$h_{ik}^{(3)}$ and $h_{ik}^{(4)}$ are traceless but have a non-zero overlap 
with ${\bf p}$ and describe the vector modes; $h_{ik}^{(5)}$ and $h_{ik}^{(6)}$ are traceless 
and have no overlap with ${\bf p}$, hence they describe the TT-tensor modes. 

Injecting this into  the linearized field equations \eqref{Eeqs},
the variables separate, giving an independent set of equations for the scalar amplitudes 
$R_1(t),R_2(t),\bm \phi(t)$, another independent set  for the vector amplitudes $R_3(t),R_4(t)$,
and finally equations for the tensor amplitudes $R_5(t),R_6(t)$. 
The scalar field perturbation $\bm \phi(t)$ contributes only to the scalar sector. 

The amplitudes $R_m$ are dimensionless but  $\bm \phi$, $\psi, \A, {\bf p}$ and the time $t$ have dimensions. 
One therefore passes 
to dimensionless variables as described by Eqs.\eqref{scale0}, \eqref{Om6}, \eqref{PSI} in the main text, so that 
$\A=\A_0 a$, etc. One also defines the dimensionless time $\tau =H_0 t$, and 
from now on the dot will denote $d/d\tau$.
The dimensionless
Hubble parameter 
\be
h=\frac{\dot{a}}{a}\equiv \frac{1}{a}\frac{da}{d\tau}
\ee
is related to $\h$ defined by Eq.\eqref{scale0} in the main text via
$h^2=\h$.  One  sets $p=\A_0 H_0 P$ and 
\be                   \label{PSI0}
\psi=-\sqrt{\frac{2\rho_{\rm cr}\Omega_6 }{3\kkappa H_0^2 } }\,\Psi,~~~~~~~~
~~~~~~
\bm \phi=-\frac{1}{H_0}\sqrt{\frac{2\rho_{\rm cr}\Omega_6 }{3\kkappa H_0^2 } }\,\phi.
\ee

As a result, 
the independent 
equations for the scalar sector amplitudes $w(\tau)\equiv R_1(\tau)$, $u(\tau)\equiv R_2(\tau)$, and $\phi(\tau)$ 
 read
\bea
3\Omega_6\Psi(\eeta-3h^2)\,\dot{\phi}&-&3h(3\Omega_6 \Psi^2+1)\,\dot{w}              \label{scal1}  \\
&-&\frac{P^2}{a^2}\left\{(\Omega_6 \Psi^2+1)(w+u)+2\Omega_6 h\Psi\,\phi\right\}=0,\nonumber  \\
(\Omega_6\Psi^2+1)(\dot{w}+\dot{u})&+&2\Omega_6 h\,\dot{\phi}          \label{scal2}
+3\Omega_6 \Psi(\eeta-h^2)\,\phi=0,                    \\
(\eeta-h^2)\,\ddot{\phi}-2h\Psi\,\ddot{w}
&+&\left(3(\eeta-3h^2)\Psi-2h\dot{\Psi}-2\Psi\dot{h}\right)\,\dot{w}     \label{scal3}  \nonumber  \\
+(3\eeta-2\dot{h}-3h^2)\,h\,\dot{\phi}  
&+&[
3(\eeta-h^2)\dot{\Psi}+9(\eeta -h^2)h\Psi)]\,w   \\
-\frac{2P^2}{3a^2}(\dot{\Psi}+h\Psi)
\,(w+u)   
&-&\frac{P^2}{3a^2}(2\dot{h}+3h^2-3\eta)\,\phi=0.  \nonumber 
\eea
Here the first two equations are first order in $d/d\tau$ because they correspond 
to the $\delta E_{00}$ and $\delta E_{03}$ gravitational 
constraints. The third equation is  second order and corresponds to $\delta E_\Phi$. 
The scalar sector contains also second order equations $\delta E_{11}$, $\delta E_{22}$, and $\delta E_{33}$,
but these are not independent and follow from \eqref{scal1}--\eqref{scal3} in view of the Bianchi 
identities. 

Equations in the tensor sector are much simpler. 
Each of the two     tensor amplitudes
 $w(\tau)\equiv R_5(\tau)$ and $u(\tau)\equiv R_6(\tau)$ fulfills exactly the same 
  equation 
  \bea
&&(\Omega_6\Psi^2+1)\,\ddot{w}+                      \label{tens}
\left(2\Omega_6 \Psi\dot{\Psi}+3(\Omega_6\Psi^2+1)h\right)\,\dot{w} \\
&&-\left(
2(\Omega_6\Psi^2+1)(2\dot{h}+3h^2)
+2\Omega_6(3\eeta\Psi^2+4h\Psi\dot{\Psi})-6\Omega_0
+\frac{P^2}{a^2}\,(\Omega_6 \Psi^2-1)
\right)w=0.  \nonumber
\eea


\providecommand{\href}[2]{#2}\begingroup\raggedright\endgroup

\end{document}